\documentclass[a4paper,11pt]{article}
\newcommand{\be}{\begin{equation}}
\newcommand{\ee}{\end{equation}}
\newcommand{\ba}{\begin{eqnarray}}
\newcommand{\ea}{\end{eqnarray}}

\usepackage{graphicx}

[1999/03/29 PSNFSS v.7.2
Times font as default roman
: S Rahtz]

\begin{document}

\title{Prediction}

\author{Didier Sornette$^1$ and Ivan Osorio$^2$\\
$^1$ ETH Zurich\\
 Department of Management, Technology and Economics\\
CH-8092 Zurich, Switzerland \\
$^2$ Department of Neurology, University of Kansas Medical Center\\ 
390 Rainbow Boulevard, Kansas City, Kansas 66160, USA}

\maketitle

\begin{abstract}

This chapter first presents a rather personal view of some
different aspects of predictability, going in crescendo from simple linear systems
to high-dimensional nonlinear systems with stochastic  forcing,
which exhibit emergent properties such as phase transitions and regime shifts.
Then, a detailed correspondence between the phenomenology of earthquakes, 
financial crashes and epileptic seizures is offered. The presented statistical evidence
provides the substance of a general phase diagram for understanding 
the many facets of the spatio-temporal organization of these systems. A key insight is to organize
the evidence and mechanisms in terms of two summarizing measures:
(i) amplitude of disorder or heterogeneity in the system and (ii) level of coupling or interaction strength
among the system's components.
On the basis of the recently identified remarkable correspondence between
earthquakes and seizures, we present detailed information on a class
of stochastic point processes that has been found to be particularly powerful
in describing earthquake phenomenology and which, we think, has a promising
future in epileptology. The so-called self-exciting
Hawkes point processes capture parsimoniously the idea that events
can trigger other events, and their cascades of interactions and mutual influence
are essential to understand the behavior of these systems.  

\end{abstract}

\noindent
{\small  chapter in ``Epilepsy: The Intersection
of Neurosciences, Mathematics, and Engineering'' ,Taylor  \& Francis Group, 
Ivan Osorio,  Mark G. Frei, Hitten Zaveri,  Susan Arthurs, eds (2010)}

\section{A brief classification of predictability}

Characterizations of the predictability (or unpredictability) of a system 
provide useful theoretical and practical measure of its complexity \cite{Boffetta02,KantzSchreiber04}.
It is also a grail in epileptology, as advanced warnings by a few minutes
may drastically improve the quality of life of these patients. 

\subsection{Predictability of linear stochastic systems}

Consider a simple dynamical system with the
following linear auto-regressive dynamics
\begin{equation}
r(t) = \beta r(t-1) + \epsilon(t)~,
\label{httwrbgqvq}
\end{equation}
where $0 < \beta < 1$ is a constant and $\epsilon(t)$ is a i.i.d. (independently identically 
distributed) random variable, i.e., a noise, with variance $ \sigma_\epsilon^2$. The dependence
structure between successive values of $r(t)$ is entirely captured by
the correlation function which is non-zero only for the time lag of one unit step (in addition
to the zero-time lag of course).
Indeed, the correlation coefficient between the random variable at 
some time and its realization at the following time step is nothing but $\beta$.
Correspondingly, the covariance of $r(t-1)$ and $r(t)$ is $\beta \times \sigma_r^2$,
where $\sigma_r^2 = \sigma_\epsilon^2 / (1-\beta^2)$ is the variance
of $r(t)$. More generally, consider an extension of expression (\ref{httwrbgqvq}) into
a linear auto-regressive process of larger order, so that we can
consider an arbitrary covariance matrix $C(t,t')$
between $r(t)$ and $r(t')$ for all possible instant pairs $t$ and $t'$.
A simple mathematical calculation shows that the best linear
predictor $m_t$ for $r(t)$ at time $t$, knowing the past history $r_{t-1},
r_{t-2}, ..., r_i,...$ is given by
\begin{equation}
m_t \equiv {1 \over B(t,t)} \sum_{i<t} B(i,t) r_i~, 
\label{formbestpred}
\end{equation}
where $B(i,t)$ is the coefficient $(i,t)$ of the inverse matrix of the covariance matrix $C(t,t')$.
This formula (\ref{formbestpred}) expresses that each past values
$r_i$ impacts on the future $r_t$ in proportion to its value with a coefficient 
$B(i,t) /B(t,t)$ which is non-zero only if there is non-zero correlation
between the realization of the variable at time $i$ and time $t$.
This formula (\ref{formbestpred}) provides the best linear predictor in the 
sense that it minimizes the errors in a variance sense. Armed with this prediction,
useful operational strategies can be developed, depending on the context.
For instance, if the set $\{r(t)\}$ denotes the returns of a financial asset,
then, one could use this prediction (\ref{formbestpred}) to invest as follows:
 buy if $m_t > 0$ (expected future price increase) and sell if $m_t < 0$ (expected future price
decrease). 

Such predictor can be applied to general moving average
and auto-regressive processes with long memory, whose general expression reads \cite{Hamiltonbook}
\begin{equation}
\left( 1 - \sum_{i=1}^p  \phi_i L^i \right)  \left(1-L\right)^d r(t) = \left( 1 + \sum_{j=1}^q  \theta_j L^j \right) \epsilon(t)~,
\end{equation}
where $L$ is the lag operator defined by $Lr(t)= r(t-1)$ and $p, q$ and $d$ can be arbitrary integers.
Such predictors are optimal or close to optimal as long as there is no change of regime, that is,
if the process is stationary and the coefficients $\{ \phi_i \}$ and $\{\theta_j\}$ and the orders
of moving average $q$, of auto-regression $p$ and of fractional derivation $d$ do not change during the
course of the dynamics. Otherwise, other methods, including Monte Carlo Markov Chains, are needed \cite{Hamiltonbook}.
In the case where the initial conditions or observations during the course of the dynamics
are obtained with noise or uncertainty, Kalman filtering and more generally data assimilation methods \cite{Ideetal97} 
provide significant improvements in predicting the dynamics of the system. 

\subsection{Predictability of low-dimensional deterministic chaotic systems}

There is an enormous amount of literature on this subject since the last 1970s
(see for instance \cite{BellaciccoKoch,KravtsovKadtke96,KantzSchreiber04,Orrell07,SmithLenny07} and references therein).
The idea of how to develop predictors for low-dimensional deterministic chaotic systems
is very natural: because of determinism, and provided that the dynamics is in some
sense sufficiently regular, the short-time evolution remembers the initial conditions, 
so that two trajectories that are found in a neighborhood of each other remain close to each other
for a time $t_f$ roughly given by the inverse of the largest Lyapunov exponent.
Thus, if one monitors past evolution, however complicated, a future path which
comes in the vicinity of a previously visited point will then evolve along a trajectory
shadowing the previous one over a time of the order of $t_f$ \cite{SugiharaMay90,Anne-Dubois91,RobinsonThiel}.
The previously recorded dynamical evolution of a domain over some short-time horizon
can thus provide in principle a short-term prediction through the knowledge of the transformation
of this domain.

However, in practice, there are many caveats to this idealized situation. Model errors
and noise, additive and/or multiplicative (also called ``parametric''), complicate and limit
predictability. Model errors refer to the generic problem that the used model
is at best only an approximation of the true dynamics, and more generally neglects
some possibly important ingredients, making prediction questionable. 

In the simplest case of additive noise decorating deterministic chaotic dynamics,
it turns out that the standard statistic methods for the estimation of the parameters
of the model break down. For instance, the application
of the maximum likelihood method to unstable nonlinear
systems distorted by noise has no mathematical ground so far \cite{PisSorchaos04}.
There are inherent difficulties in the statistical analysis
of deterministically chaotic time series due to 
the tradeoff between the need of using a large number of data points in the maximum
likelihood analysis to decrease the bias and to
guarantee consistency of the estimation, on the one hand, and the unstable nature of dynamical trajectories with exponentially
fast loss of memory of the initial condition, on the other hand. The method of statistical moments for the estimation of the
parameter seems to be the unique method whose consistency for
deterministically chaotic time series is proved so far theoretically (and not just numerically)  \cite{PisSorchaos04}.
But the method of moments is well-known to be relatively inefficient.

\subsection{Predictability of systems with multiplicative noise}

The presence of multiplicative (or parametric) noise makes the dynamics much richer... and
complex. New phenomena appear, such as stochastic resonance \cite{GHJM:1998},
coherence resonance \cite{PK:1997}, noise-induced
phase transitions  \cite{HorsthemkeLefever83,SanchoOjalvo2000}, noise-induced transport 
 \cite{HM:2009} and its game theoretical version, the Parrondo's Paradox \cite{Abbott2010Asymmetry}.
 The predictability is then a non-monotonous
function of the noise level. Even the simplest possible combination of nonlinearity
and noise can utterly transforms the nature of predictability. Consider for instance the bilinear 
stochastic dynamical system, arguably the simplest incarnation of nonlinearity (via bilinear
dependence on the noise) and stochasticity:
\begin{equation}
r(t) = \epsilon(t) + b \epsilon(t-1) \epsilon(t-2)~,
\label{hty5h2yt}
\end{equation}
where $\epsilon(t)$ is a i.i.d. noise. The dynamics (\ref{hty5h2yt}) is the simplest
implementation of the general Volterra discrete series of the type
\begin{equation}
r(t) = H_1\left[\epsilon(t)\right]  + H_2\left[\epsilon(t)\right]  + H_3\left[\epsilon(t)\right] + ... + H_n\left[\epsilon(t)\right] + ...
\end{equation}
where
\begin{equation}
H_n\left[\epsilon(t)\right]  = \sum{j_1=0}^{+\infty} ... \sum{j_n=0}^{+\infty} h_n(j_1, ..., j_n) \epsilon(t-j_1) ... \epsilon(t-j_n)~.
\end{equation}
By construction, the time series $\{r(t)\}$ generated by
expression (\ref{hty5h2yt}) has no linear predictability (zero two-point
correlation) but a certain nonlinear predictability (non-zero
three-point correlation) \cite{PisSorchaos08}. It can thus be considered as a paradigm for
testing the existence of a possible nonlinear predictability in a given
time series. Notwithstanding its remarkable simplicity, the
bilinear stochastic process (\ref{hty5h2yt}) exhibits remarkably rich and complex
behavior. In particular, the inversion of the key
nonlinear parameter $b$ and of the two initial conditions necessary for the
implementation of a prediction scheme exhibits a quite anomalous 
instability: in the presence of a some random large impulse of the exogenous noise
$\epsilon(t)$, the ensuing dynamics exhibits super-exponential sensitivity 
for the inversion of the innovations  \cite{PisSorchaos08}.

\subsection{Higher dimensions, coherent flows and predictability}

Going bottom-up in the complexity hierarchy, we have
low-dimensional chaos $\to$ spatio-temporal chaos \cite{CrossHohenberg93} $\to$ turbulence \cite{UrielFrisch}.
It turns out that, contrary to naive expectation, increasing dimensionality and introducing
spatial interactions does not necessary destroy predictability. This is due to the organization
of the spatio-temporal dynamics in so-called ``coherent structures'', corresponding to coherent vortices
in hydrodynamic flows \cite{Boffettacrisanti97}. It has been shown that
 the full nonlinearity acting on a large number of degrees of freedom can, paradoxically, 
 improve the predictability of the large scale motion, 
 giving a picture opposite to the one largely popularized by Lorenz for 
 low-dimensional chaos. The mechanism for improved predictability is that 
small local perturbations can progressively grow to larger and larger scales by nonlinear interaction 
and finally cause macroscopic organized persistent structures \cite{RobertRosier01}.

\subsection{Fundamental limits of predictability and the virtue of coarse-graining}

Algorithmic information theory\index{Algorithmic information theory} \cite{LiVitanyi} combines information theory, computer science and meta-mathematic logic.
In the context of system predictability, it has profound implications.  Indeed, 
a central result of algorithmic information theory obtained as a synthesis of the efforts
of R. Solomonoff \cite{Solomonoff09}, A. Kolmogorov, G. Chaitin \cite{Chaitin87}, P. Martin-L\"of, M. Burgin and others states
roughly that ``most'' dynamical systems evolve according to and/or produce outputs that
are utterly unpredictable. Here, the term ``most'' in ``most dynamical systems'' mean that this 
property holds with probability $1$ when choosing at random a dynamical system from 
the space of all possible dynamical systems.
Specifically, the data series produced by most dynamical systems have been
proved to be computationally irreducible, i.e. the only way to decide about their evolution is 
to actually let them evolve in time. There is no way you can compress their dynamics
and the resulting information into generation rules or algorithms that are shorter than 
the output itself. Then, the only strategy is to let the system evolve and reveal its
complexity, without any hope of predicting or characterizing in advance its properties.
 The future time evolution of most complex systems thus appears inherently unpredictable. 
This is the foundation for the approach pioneered by S. Wolfram \cite{Wolfram02} to basically renounce the hope
to get mathematical laws and predictability, and replace them by the search for
cellular automata\index{cellular automata} that have universal computational abilities (like so-called Turing machines)
and can reproduce any desired pattern. 

Such views are almost shocking to most scientists, whose job is to find patterns that can be 
captured in coherent models that provide a reduced encoding of the observed complexity,
in direct apparent contradiction with the central result of algorithmic information theory.
Israeli and Goldenfeld have provided an insightful and elegant procedure, based
on renormalization group theory, to reconcile the two view points \cite{IsraeliGoldenfeld04,IsraeliGoldenfeld06}.
The key idea is to ask only for approximate answers, which for instance makes
physics work, unhampered by computational irreducibility. 
By adopting the appropriate ``coarse-grained''\index{coarse-graining} perspective of how to study the system,
Israeli and Goldenfeld found that even the known computational irreducible cellular
automaton (rule 110 in Wolfram's classification \cite{Wolfram83}) becomes relatively simple and predictable. 
In physics, this comes as no surprise. Each trajectory of the approximately $10^{25}$ molecules
in an office room each follows an utterly chaotic trajectory, which loses predictability
after a few inter-molecular collisions. But the coarse-grained large-scale properties
of the gas is well-captured by the law of ideal gas $pV = n RT$, or van der Waals' equation
if one wants a bit more precision, where $p$ is the pressure in the enclosure of volume $V$
at temperature $T$, and $n$ is the number of moles of gas, while R is a constant.
Thus, but asking questions involving different scales, 
computationally irreducible systems can be predictable at some level of description.
The challenge is to find how to coarse-grain, what is the optimal level of
description, and what effective macroscopic interactions and patterns emerge
from this procedure. There are promising developments in this direction 
to elaborate a general theory of hierarchical dynamics \cite{NilssoJacobin1,NilssoJacobin2,NilssoJacobin3},
using the renormalization group as a constructive meta-theory of model building \cite{Wilson79}.

\subsection{``Dragon-kings''}

Predictability may come from another source, that is, directly from specific
transient structures developing in the system, that we refer to
as ``dragon-kings''\index{dragon-king} \cite{Dragonkings09,SatinoverSornette10}.

The concept of dragon-kings has been introduced as a frontal refutal
to the claim that ``black swans''\index{black swan} 
characterize the dynamics of most systems \cite{Talebswan07}. 
According to the ``black swan'' hypothesis,
highly improbable events with extreme sizes
or impacts are thought to occur randomly, without any precursory
signatures. ``Black swan'' events are thought to be events
of large sizes associated with the tail of distributions such as
power laws. Because the same power law distribution 
is thought to describe the whole population of event sizes, including
the ``black swans'' of great impact, the argument is that there
are not distinguishing features for these ``black swans'', except their
great sizes, and therefore no way to diagnose their occurrence in 
advance. In this story, for instance, a great earthquake is just an event that 
started as a small earthquake... and did not stop growing. Its
occurrence is argued to be inherently unpredictable because
there is not way to distinguish the nucleation of the myriads
of small events from the rare ones that will grow to great sizes
by chance \cite{Gelleretal97}. 

In contrast, the ``dragon-king'' hypothesis proposes that 
extreme events from many seemingly unrelated domains may be plausibly 
understood as part of a different population than that comprising the large
majority of events. This difference may result from amplifying mechanisms,
such as positive feedbacks, which are active only transiently, leading to the emergence of
non-stationarity structures. The term ``dragon'' refers to the mythical animal
that belongs to a different animal kingdom beyond the normal, with extraordinary characteristics.
The term ``king'' had been introduced earlier \cite{LahSor98}
to emphasize the importance of those events, which are beyond the extrapolation of 
the fat tail distribution of the rest of the population. This is in analogy with the sometimes 
special position of the fortune of kings, which appear to exist beyond the Pareto
law distribution of wealth of their subjects  \cite{Dragonkings09}.
The concept of dragon-kings has been argued to be relevant under a broad
range of conditions in a large variety of systems, including the distribution of city
sizes in certain countries such as France and the United Kingdom, 
the distribution of acoustic emissions associated with material
failure, the distribution of velocity increments in hydrodynamic turbulence,
the distribution of financial drawdowns, the distribution of the energies of
epileptic seizures in humans and in model animals, the distribution of the
earthquake energies and the distribution of avalanches in slowly driven 
systems with frozen heterogeneities (see \cite{Dragonkings09,SatinoverSornette10}
for a detailed presentation of these various examples and the related bibliography).

\subsection{A Landau-Ginzburg\index{Landau-Ginzburg} model of self-organized critical avalanches coexisting with Dragon-kings
\label{sectLG}}

The following model \cite{Gil-Sornette96} provides a quite generic set-up for the emergence
of dragon-kings under wide parameter conditions, coexisting with a self-organized
critical regime under different parameter conditions. This model 
is relevant to a large number of systems, 
including systems of coupled neurons. Consider an extended system, 
whose local state at position ${\vec r}$ and time $t$
is characterized by the local order parameters $S({\vec r}, t)$.
The order parameter $S$ is zero in absence of activity
and non-zero otherwise. Its amplitude $S({\vec r}, t)$ 
quantifies the level of activity at ${\vec r}$ and time $t$. 

\begin{figure}
\includegraphics[width=350pt]{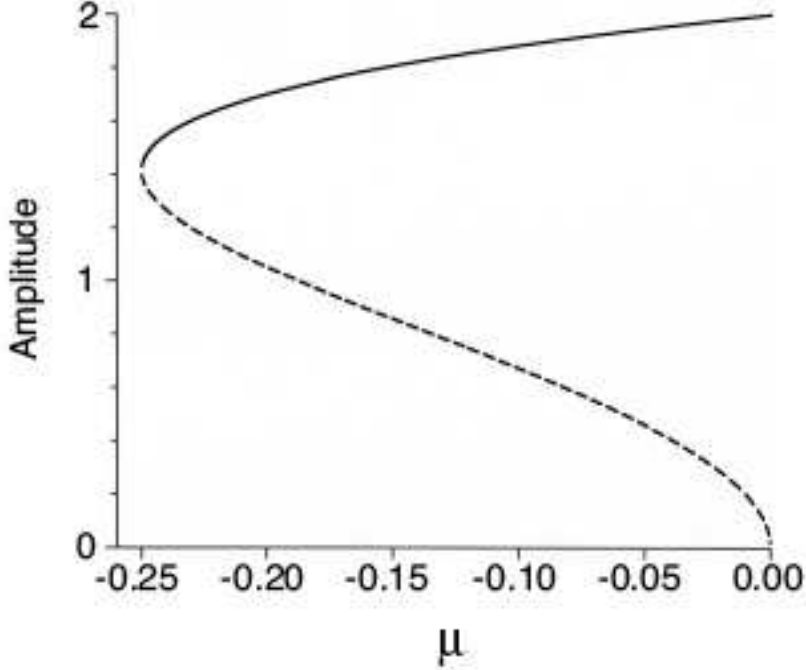}
\caption{Bifurcation diagram in the positive domain $S>0$ of the normal form (\ref{t3yhybqv})
plotting the amplitude $|S|$ as a function of the control parameter $\mu$.
}
\label{Fig_bifurcation-sub}
\end{figure}

The simple and general dynamical
equation that captures the process of jumps between a zero 
to a non-zero activity state consists of the normal form of the
sub-critical pitchfork bifurcation\index{bifurcation} of co-dimension $1$:
\begin{equation}
{\partial S \over \partial t} = \chi \left( \mu S + 2 \beta S^3 - S^5 \right)~.
\label{t3yhybqv}
\end{equation}
The parameter $\chi$ sets the characteristic time scale $1/\chi$ of the 
dynamics of $S$.
The parameter $\beta$ is taken positive, corresponding to the sub-critical pitchfork
bifurcation regime. In absence of the stabilizing $- S^5$ term, the non-zero
fixed points (for the relevant regime $\mu<0$) given by $S^*_{\pm} = \pm \sqrt{-\mu / 2\beta}$ are unstable,
while the fixed point $S_0=0$ is locally stable. These two unstable fixed points 
correspond to the dashed line in figure \ref{Fig_bifurcation-sub}.
The term ``locally'' reflects
the fact that a sufficiently large perturbation that pushes $S$ above $S^*_+$
or below $S^*_-$ will be amplified leading to a diverging amplitude $|S|$ at long times.
In the presence of the $- S^5$ term, two new fixed point exist, which are stable.
They correspond to the upper solid line in figure \ref{Fig_bifurcation-sub}.
The bifurcation diagram of these fixed points as a function of $\mu$ shown in figure
\ref{Fig_bifurcation-sub} is similar to the bifurcation diagram of the Hodgkin-Huxley model, 
for which the transmembrane voltage is the order parameter $S$ and the external
potassium concentration is the control parameter $\mu$.

Now, imagine that the normal form (\ref{t3yhybqv}) describes the local state $S({\vec r}, t)$
at  ${\vec r}$ and time $t$, which may be different from point to point because
the control parameter $\mu$ is actually dependent on position ${\vec r}$ and time $t$.
We thus have as many dynamical equations of the form (\ref{t3yhybqv}) as there 
are points ${\vec r}$ in the system. For each point ${\vec r}$, the local control parameter
$\mu({\vec r}, t)$ is assumed to be an affine function of
the gradient of a local concentration $h$:
\begin{equation}
\mu(r, t) = g_c - {\partial h \over \partial r}~.
\label{yhyth2tg}
\end{equation}
We consider a cylindrical (or one-dimensional) geometry so that a single
spatial coordinate $r$ is sufficient (and we can drop the arrow
on ${\vec r}$).  Here, $g_c$ is
the critical value of the gradient at which the zero-fixed point 
$S_0=0$ becomes linearly unstable. The model in Ref.~\cite{Gil-Sornette96} 
assumed a slightly different technical form ($\mu(r, t) = g_c - \left({\partial h \over \partial r}\right)^2$),
which does not change the main regimes and results described below.

\begin{figure}
\includegraphics[width=350pt]{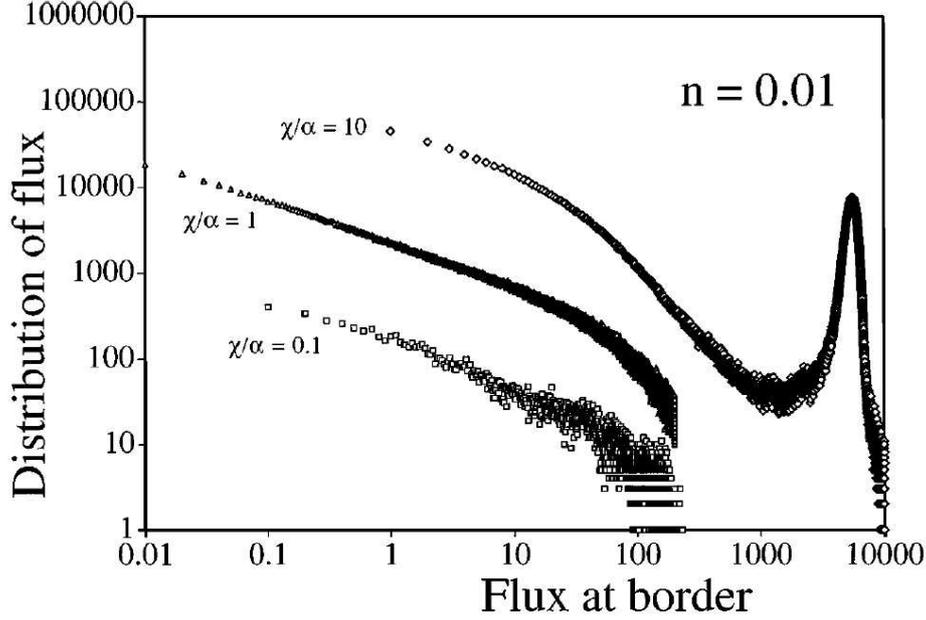}
\caption{Distribution of flux amplitudes at the open border of the one-dimensional system
obeying the dynamics described by expressions (\ref{t3yhybqv},\ref{yhyth2tg},\ref{thuyj6ujujk},\ref{tjuyjkuk57k}).
The standard deviation $n$ of the noise term $n(r,t)$ is equal to $0.01$ (small driving noise regime).
Reproduced from Ref.~\cite{Gil-Sornette96}.}
\label{Figflux-Landau}
\end{figure}

Because we think of $h(r, t)$ as a diffusing field, its equation of evolution is generically
\begin{equation}
{\partial h \over \partial t} = - {\partial F\left(S, {\partial h \over r}\right) \over \partial r} + n(r,t)~.
\label{thuyj6ujujk}
\end{equation}
This equation expresses that the rate of change of $h$ is equal to the gradient of a flux
that ensures the conservation of the concentration, up to an external fluctuation noise
$n(r,t)$ acting on the system. The last ingredient of the model consists in writing that
the flux is proportional to the gradient of the field:
\begin{equation}
F\left(S, {\partial h \over r}\right) = - \alpha ~S^2~{\partial h \over \partial r}~,
\label{tjuyjkuk57k}
\end{equation}
where $\alpha$ is another inverse time scale controlling the diffusion rate of the field
within the system. The proportionality between the flux $F$ and the gradient 
$-{\partial h \over \partial r}$ of the field is simply Fick's law. The non-standard ingredient
stems from the fact that the coefficient of proportionality, usual defining the diffusion
coefficient, is controlled by the amplitude $S^2$ of the order parameter. 
In absence of activity $S=0$, the local flux $F$ is here zero and the field does not change, up
to noise perturbations. This corresponds to a
strong feedback of the order parameter onto the control parameter,
which has been shown to be one of the possible mechanism for the emergence
of self-organized criticality\index{self-organized criticality} \cite{SornetteSOC92,Fraysseetal93,Gil-Sornette96}.
Recall that standard formulations of the dynamics
and bifurcation patterns of evolving systems in terms of normal forms assume the existence of control
parameters that are exogenously determined. Here, the order parameter
of the dynamics has an essential role in determining the value of the control parameter, 
which becomes itself an endogenous variable. 

The analysis of the dynamics described by expressions (\ref{t3yhybqv},\ref{yhyth2tg},\ref{thuyj6ujujk},\ref{tjuyjkuk57k})
presented in Ref.~\cite{Gil-Sornette96} shows that a self-organized critical (SOC) regime \cite{Bak96} appears
under the condition of small driving noise and when the diffusive relaxation is faster than the instability growth
rate:  $\alpha > \chi$. The SOC dynamics can be shown to be associated with a renormalized
diffusion equation at large scale with an effective negative diffusion coefficient \cite{Gil-Sornette96},
expressing that small scale fluctuations are the most unstable and cascade intermittently to large 
scale avalanches. This SOC regime is exemplified by the power law distributions of avalanche sizes
shown in figure \ref{Figflux-Landau} for  $\chi/\alpha =0.1$ and $1$.
More interesting for our purpose is the fact that, when $\alpha < \chi$, characteristic
large scale events appear, which coexist with a crowd of smaller events themselves approximately 
distributed according to a power law with an exponent larger than in the SOC regime.
The dragon-kings\index{dragon-king} correspond to the peak on the right of figure \ref{Figflux-Landau}, associated
with the run-away avalanches of size comparable to the size of the system.
 
This constitutes an example of what we believe to be a generic behavior 
found in systems made of heterogeneous coupled threshold oscillators\index{threshold oscillators}, such as sandpile models, 
Burridge-Knopoff block-spring models \cite{Schmittbuhletal93} and earthquake-fault models 
\cite{Soretal94,SorMilVan95,Dahmenetal98}: 
a power law regime (self-organized critical) (Figure \ref{fig9_diagram}, right lower half) is 
co-extensive with one of synchronization \cite{strogatz04} 
with characteristic size events (Figure \ref{fig9_diagram}, upper left half). 
We discuss below this generic phase diagram, in our attempt to 
compare the dynamics and resulting statistical regularities observed in earthquakes,
financial fluctuations and epileptic seizures.

\subsection{Bifurcations, Dragon-kings and predictability \label{sectBifDragon}}

The existence of ``dragon-kings'' punctuating the dynamics
of a given system suggests mechanisms of self-organization otherwise not
apparent in the distribution of their smaller siblings. Therefore, 
this opens the potential for predictability, based on the hypothesis
that these specific mechanisms that are at the origin of the dragon-kings could leave
precursory fingerprints usable for forecasts.

The dynamical system (\ref{t3yhybqv},\ref{yhyth2tg},\ref{thuyj6ujujk},\ref{tjuyjkuk57k})
presented in the previous section shows an example in which 
the dragon-kings appear in a large range of parameters in the presence
small scale subcritical bifurcation dynamics\index{bifurcation}, which are renormalized
at large scales into a change of regime, a bifurcation of behavior, 
more generally a transition of phase. In other words, 
dragon-kings are commonly associated with a phase transition. If a phase transition can
be detected before it occurs, it may be understood as an abrupt
increase in the probability, or risk, of an extreme event. Practical
examples include ruptures in materials and bursting of financial bubbles.

Mathematicians have proved \cite{Thom,Arnold} that,
under fairly general conditions, the local study of  bifurcations 
of almost arbitrarily complex dynamical systems can be
reduced to a few archetypes. More precisely, it is proved that there exists
reduction processes, series expansions and changes of variables of the many
complex microscopic equations such that, near the fixed point (i.e. for 
small values of the order parameter $S$),
the behavior is described by a small number of ordinary differential
equations depending only a few control parameters, like $\mu$ in expression (\ref{t3yhybqv})
for a sub-critical pitchfork bifurcation. The result is
non-trivial since a few  effective numbers such as $\mu$
 represent the values of the various relevant control variables and a single
(or just a few) order parameter(s) is(are)
sufficient to analyze the bifurcation instability. The remarkable consequence
is that the dynamics of the system in the vicinity of the bifurcation is 
reducible and thus predictable to some degree. This situation can be described
as a reduction of dimensionality or of complexity, that occurs
in the vicinity of the bifurcation. Such reduction of complexity may occur
dynamically and intermittently in large dimensional out-of-equilibrium systems, 
such as in hierarchically coupled Lorenz systems \cite{Lozenz-hierarchy} or in agent-based models
of financial markets \cite{AndersenSornette05}.

As an illustration, consider expression (\ref{t3yhybqv}) where $\beta$ is now assumed
negative. Since the cubic term $2 \beta S^3$ is now stabilizing, the quintic term $- S^5$ can be dropped.
An interesting so-called super-critical bifurcation occurs at $\mu=0$,
separating the regime for $\mu<0$ where the zero-fixed point $S_0=0$ is unique and
is stable, from the regime $\mu>0$ where two symmetric stable fixed points
appear at $S^*_{\pm} = \pm \sqrt{\mu / 2|\beta|}$, and the zero-fixed point $S_0=0$
becomes unstable. Consider the dynamics of such a system slightly perturbed by
an external noise $n(t)$ with zero mean and variable $\sigma^2$, so that its dynamics reads
\begin{equation}
{dS \over dt} =  \mu S - 2 |\beta| S^3 + n(t)  ~.
\label{t3yhybqrggv}
\end{equation}
For $\mu <0$, the average value $\langle S(t) \rangle$ vanishes but its variance
can be calculated explicitly from the solution of (\ref{t3yhybqrggv}). Indeed, to 
a very approximation, we can drop also the $2 |\beta| S^3$ term since $S$ is 
exhibiting only small fluctuating excursions around $0$, for $\mu <0$, according to 
\begin{equation}
S(t) = \int_{-\infty}^t e^{-|\mu| (t-\tau)}~ n(\tau) d\tau~.
\end{equation}
Its variance  $\langle [S(t)]^2 \rangle$ is then given by
\begin{equation}
 \langle [S(t)]^2 \rangle = \sigma^2  \int_{-\infty}^t e^{-2|\mu| (t-\tau)} d\tau  = {\sigma^2  \over 2 |\mu|}~.
 \label{hy3uj6k7i}
 \end{equation}
 This result (\ref{hy3uj6k7i}) shows that the variance $\langle [S(t)]^2 \rangle$ 
 of the fluctuations of the order parameter
 diverges as the critical bifurcation point is approached from below: $\mu \to 0^-$. 
 $\langle [S(t)]^2 \rangle$ plays the role of a susceptibility, whose divergence
 on the approach to the critical point suggests a general predictability, for instance
 obtained by monitoring the growth and correlation properties
 of the system fluctuations. This method has been
used in particular for material failure (recording of micro-damage
for instance via acoustic emissions) \cite{Anifrani,Ciliberto,Critrup00}, human parturition
(proposed recording of the mother-foetus maturation process via Braxton-hicks contractions of the uterus)
\cite{Parturition1,Parturition2}, financial crashes (monitoring of bursts
of price acceleration and various risk measures via options)  \cite{crash1,riskpaper,JSL,Sornettewhystock03}
and earthquakes (monitoring of precursory seismic, electromagnitic
and chemical activity)  \cite{seismic,Johansekobe,Bowman}.
We believe that this phase transition approach bears great potential 
to predict catastrophic events, recognizing precursors
in time series associated with finite-time singularities \cite{JohSorfts01,SammisSornette02,IdeSorfts02,GluzmanSorfts02},
hierarchical power law precursors \cite{SorPredictPNAS02}, critical slowing down
\cite{Dakosetal08} and other types of precursors \cite{sornette06,Schefferetal09}.

\section{Parallels between earthquakes, financial crashes and epileptic seizures}\label{sesmic-finance-seizure}

How can the concepts described in the previous
section be applied to real systems, and in particular to the prediction of epileptic seizures?
To put this question in a broader perspective, we present in this section an original attempt 
\cite{Osorioetal1, Osorioetal2}
to draw parallels between seemingly drastically different systems and phenomena, based
on both qualitative and quantitative evidence. 

\subsection{Introduction to earthquakes, financial crashes and epileptic seizures}

\begin{figure}
\includegraphics[width=350pt]{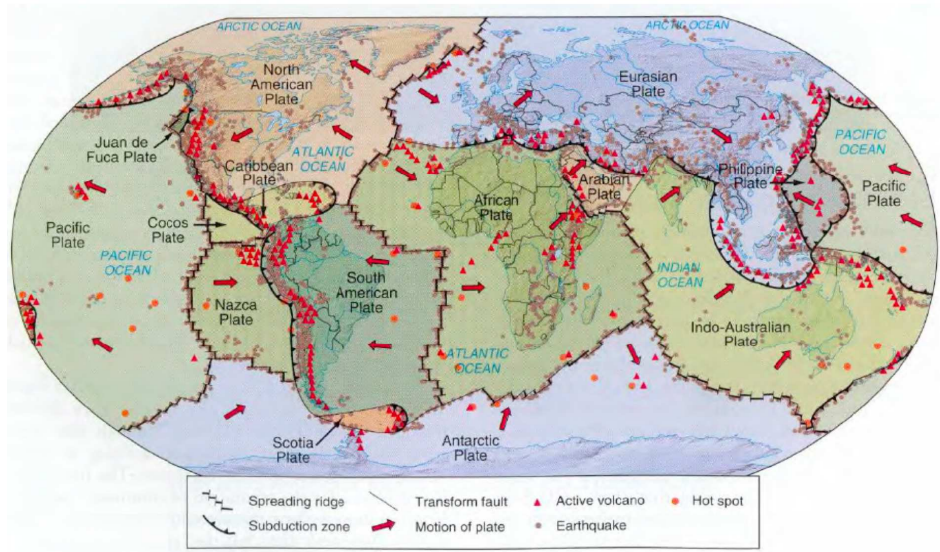}
\caption{Standard model of 12 major tectonic plates showing their relative motions (thick red arrows), and
the plate boundaries which concentrate a large fraction of seismic and volcanic activity.  The 
three types of the place boundaries are also indicated in the legend.}
\label{fig1_plates}
\end{figure}

\begin{figure}
\includegraphics[width=350pt]{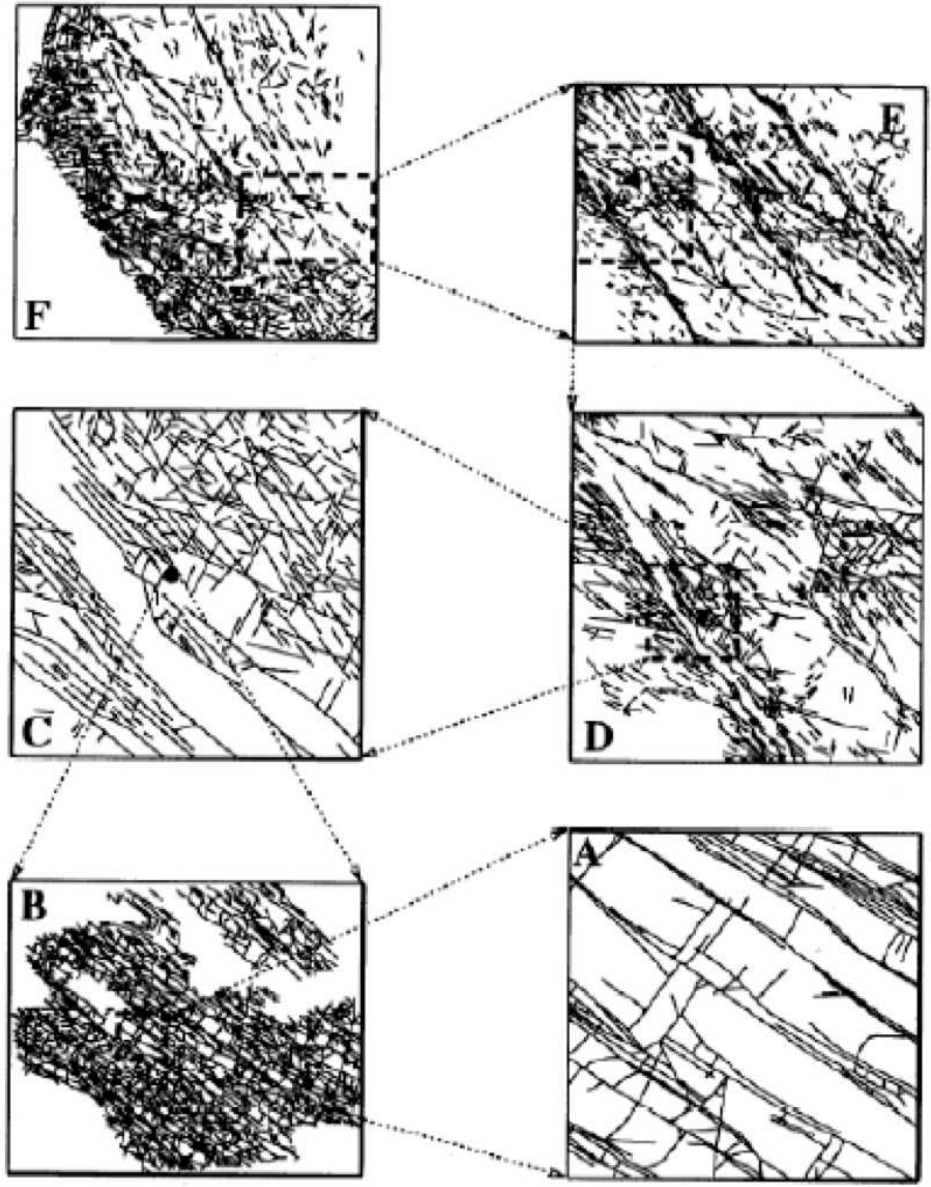}
\caption{Example illustrating the hierarchical organization of faults from the 400 km scale (upper left panel)
down to the 1 meter scale (lower right panel). Reproduced from ref.~\cite{Ouillonetal96}.}
\label{fig2_hierar-faults}
\end{figure}

Earthquakes occur mostly in the thin upper fragile layer of the Earth, called the upper crust. 
A complex system of slowly moving tectonic plate boundaries delineate their most probable location,
as shown in figure \ref{fig1_plates}.
Recent syntheses of compendium of geological and seismic data \cite{Bird03} suggest that the system of tectonic plates
covering the Earth surface is ``fractal'' \cite{SorPis03}, i.e., composed of a broad (power law) distribution of plates.
Even more interesting is the fact that, in broad region around the
tectonic plate boundaries, earthquakes are clustered on networks of faults forming
rich hierarchical structures from the thousand kilometer scale
to the meter scale and below \cite{Ouillonetal96}, as shown in figure \ref{fig2_hierar-faults}. 
At a qualitative level (and supported quantitatively by some models \cite{Soretal94,Lyakhovskyetal01}), 
it is thought that the fault networks are self-organized by the repetitive action of earthquakes.

Financial crashes occur in organized markets trading assets, such as equities of firms, commodities
such as oil or gold, and bonds (debts of firms or of countries). By their varying
and heterogeneous demand and supply, investors
are responsible for the observed price variations. Investors come in a very broad
distribution of sizes (and therefore market impacts), from the individual private household
to the largest pension and mutual funds, commanding up to hundreds of billions
of dollars. These investors are interacting with other investors as well as with market makers,
with commercial and investment banks, as well as more recently with sovereign funds. 
This variety of sizes, needs and goals provides a fertile ground for rich behaviors, including
systemic instabilities and crippling crashes.

\begin{figure}
\includegraphics[width=350pt]{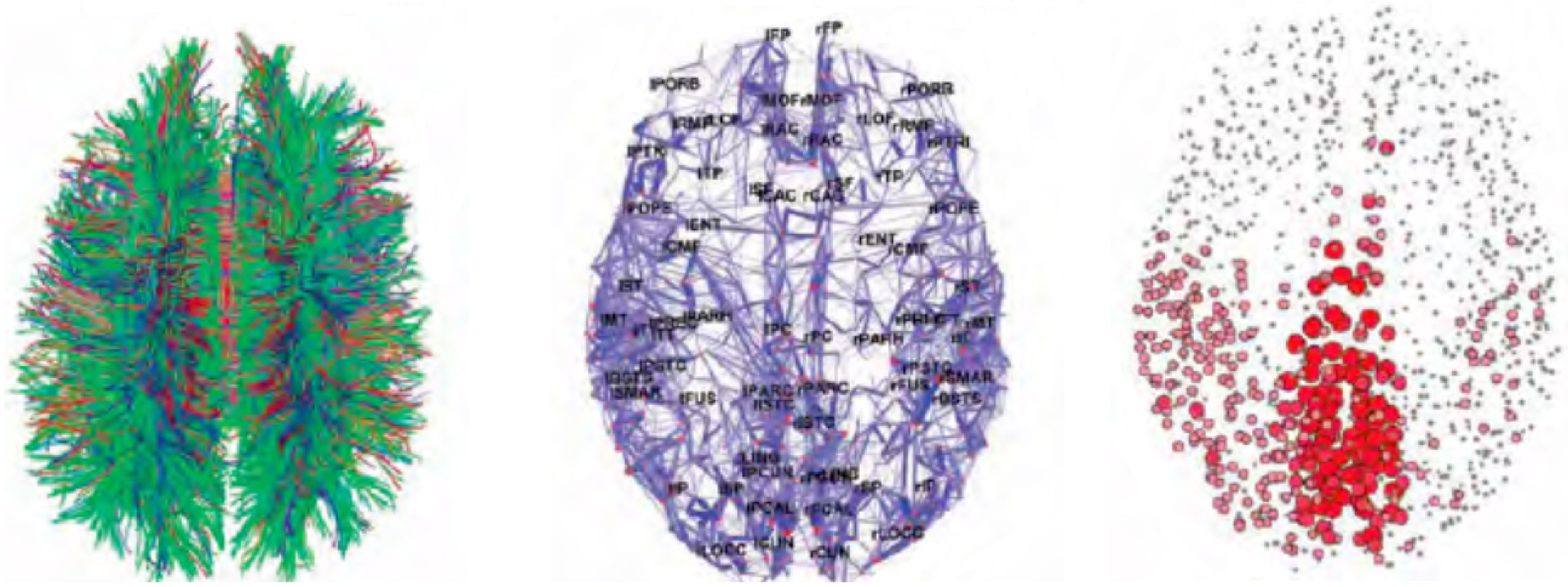}
\caption{Illustration of the complex hierarchical network structure of the brain.
Left: fiber pathways of the human cerebral cortex; middle:  network of connections in the human
cortex, with lines between brain regions indicating the strengths of the connections;
right: location of highly connected hub nodes forming the structural core.
Reproduced from  Ref.~\cite{Hagmannetal08}.}
\label{fig3-brain-hier}
\end{figure}

Epileptic seizures occur within what many refer exaggeratedly to as the most complex system
of this universe, the human brain. The human brain is organized in an exceedingly rich
set of topographic and functional divisions at many scales, from the lobes and
complex folded structures down to columns and
to neurons (see figure \ref{fig3-brain-hier} for a partial insight in this
rich organization).  The networking and function of these units reflect both encoded 
development programs as well as the impact of learning and experience that feedback
on the development processes.

\subsection{Common properties between earthquakes, financial crashes and epileptic seizures \label{tyjrukjiuk}}

\begin{figure}
\includegraphics[width=350pt]{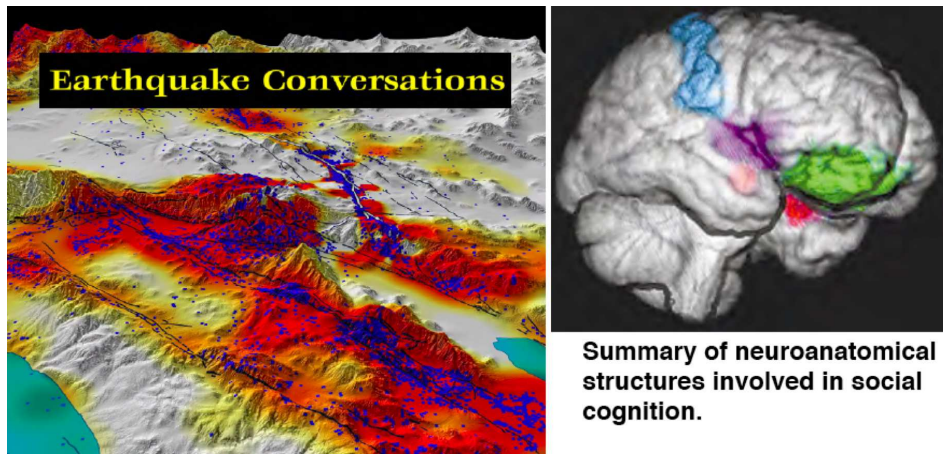}
\caption{Left panel: cover picture of the Scientific American journal in with R.S. Stein \cite{Stein03} reviews
the evidence for ``earthquake conversations'', that is, the predominant hypothesis
that earthquakes trigger earthquakes. The right panel outlines several important
structures involved in social cognition and interactions 
(ventromedial prefrontal cortex (green), amygdala (red), right somatosensory cortex (blue)
and insula (purple)). Reproduced from  Ref.~\cite{Hagmannetal08}.}
\label{fig4_conversations}
\end{figure}

Earthquakes, financial crashes and epileptic seizures are characterized by several strikingly similar 
mechanisms and properties.
\begin{enumerate}
\item They occur on  hierarchically organized structures, with many inter-connected scales.
\item Their distribution in sizes are heavy tailed and extreme events are typical.
\item There is a strong entanglement between the growth and properties of the supporting structures
and the spatio-temporal organization of the events themselves: 
the supporting structures and the events inter-organize as in a chicken-and-egg problem:
earthquakes occur on faults and faults grow and form networks shaped by the repetition
of earthquakes; financial crashes occur on financial markets acted by investors
whose actions and impacts result from the cumulative growth of their fortunes shaped
by past financial performance, which feedbacks on future performance. Young brains
grow with epileptic regimes (e.g., ``absence'' seizures) and there are many feedbacks
between structures and functions. This suggests that a genuine understanding of
the generating processes and of the properties 
of earthquakes, financial crashes and epileptic seizures can only be obtained by
studying the joint organization of these events and their evolving self-organized
carrying structures. The basis (bases) for this important statement is (are) not well
developed for seizures.
\item Past earthquakes trigger\index{triggering} future earthquakes: it is estimated that between 50\% to close
to 100\% of earthquakes are triggered by past earthquakes (and not just the aftershocks).
This is illustrated by the concept that earthquakes have ``conversations'',
similarly to the exchanges between different areas of the brain when developing
cognitive tasks (see figure \ref{fig4_conversations}). 
Most of the volatility of financial markets is probably the result of endogenous 
amplification of past returns on future returns rather than the direct exogenous
effect of external news, as for instance exemplified by the so-called ``excess volatility'' effect.
The concept that seizures beget seizures has a long history and new recent 
empirical evidence supports the rational to revisit this hypothesis \cite{Osorioetal1,Osorioetal2}.
\item Within a coarse-grained\index{coarse-graining} approach to the modeling of these systems, they
can be represented as made of coupled threshold oscillators\index{threshold oscillators} of relaxation
(faults going to rupture, investors going to investment decisions, neurons going to a
firing state).
\item There is some evidence that these three systems are characterized by the coexistence
of scaling (power law distribution of event sizes) and regimes with large
characteristic events\index{dragon-king} \cite{Dragonkings09}.
\item Finally, there is a lot of interest in our modern societies to diagnose and predict large 
catastrophic events, to help alleviate the damage associated with earthquakes, the losses of financial crashes
and to help patients recover normal lives in the presence of intermittent seizures.
\end{enumerate}

\begin{figure}
\includegraphics[width=350pt]{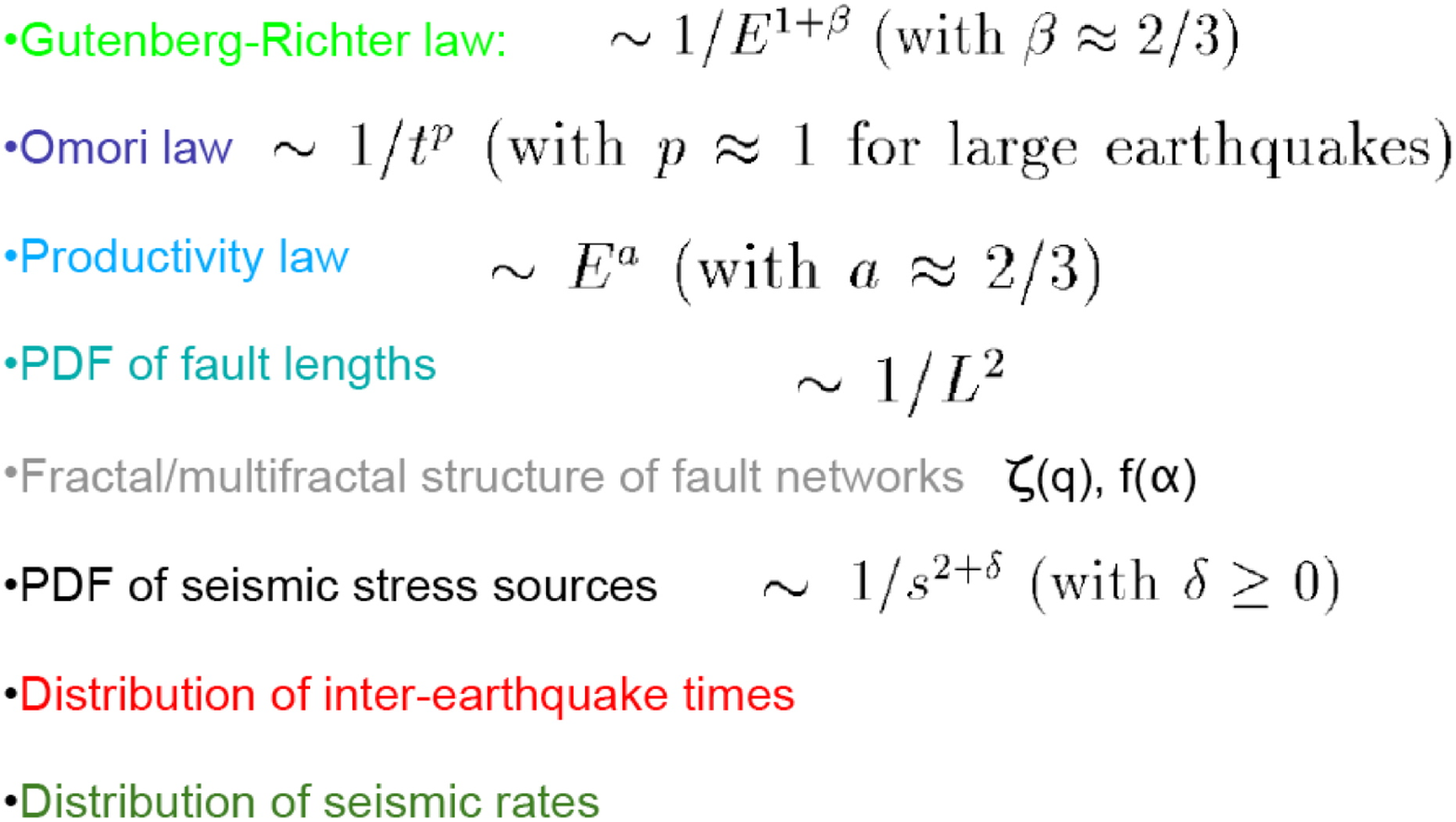}
\caption{Survey of the major statistical laws in seismicity. The color code allows for comparison
with the statistical laws in finance markets and in epileptic seizures documented in figures 
\ref{fig6_statlawfinance} and \ref{fig7_statlawseizures}.}
\label{fig5_statlawEQ}
\end{figure}

Figure \ref{fig5_statlawEQ} summarizes the main statistical laws that have been documented
in seismology  (see Ref.~\cite{SornetteGeilo} and references therein). 
\begin{enumerate}
\item The Gutenberg-Richter\index{Gutenberg-Richter law} law describes the probability density function (pdf)
of earthquakes of a given energy $E$, as being a power law with a small exponent $\beta \approx 2/3$.
\item The Omori law\index{Omori law} states that the rate of aftershocks\index{aftershocks} following an earthquake (usually
improperly referred to as a ``main shock'') exhibits a burst immediately after
the main shock and decays slowly in time afterwards as the inverse of time raised
to an exponent $p$, which is close to $1$ for large earthquakes. 
\item The productivity law
describes how the average number of triggered earthquakes depend on the energy $E$ of
the triggering earthquake: the larger an earthquake, the more earthquakes it triggers, 
according to a power law with an exponent $a$ probably slightly smaller than $\beta$ \cite{Helmstetter03}.
\item Because earthquakes occur on faults, and faults grow by earthquakes, it is important
also to characterize the properties of fault networks. It is well-documented
that the probability density distribution $P(L)$ of fault lengths in a given area is described by
a power law $P(L) \simeq 1/L^f$ with exponent $f$ not far from $2$.
\item Several studies have documented that fault networks exhibit fractal, multifractal or
better multi-scale hierarchical properties \cite{Ouillonetal96}.   
\item Earthquakes
result from deformations that produce complex stress fields, which are one of the 
important fields at the origin of the nucleation of earthquakes (the distribution
of water (brine) in the crust is also thought to play a crucial role, albeit we have
only indirect and incomplete information, see Refs.~\cite{Sornettemechano1,Sornettemechano2}
for a review). The distribution of stress amplitudes have been documented
from the focal source mechanism of earthquakes to be close to a Cauchy 
distribution, i.e., with a power law tail $\simeq 1/s^{2 +\delta}$ and $\delta$ small \cite{Kaganstress94}.
\item The distribution of waiting times between earthquakes in a given region is also
characterized by a fat tail, approximately quantified by a power law,
indicative of a broad range of inter-event intervals. However, recent studies 
suggest that the pdf of inter-earthquake intervals has several regimes 
(see Ref.~\cite{SaichevSorPRL04,SaiSor07,SorUtkinSai08} and references therein) 
and may not be describable by a simple power law.  
\item The distribution of seismic rates
(number of earthquakes per unit time) in fixed regions is also well-described by a power law function \cite{SaiSor06}.
\end{enumerate}

\begin{figure}
\includegraphics[width=350pt]{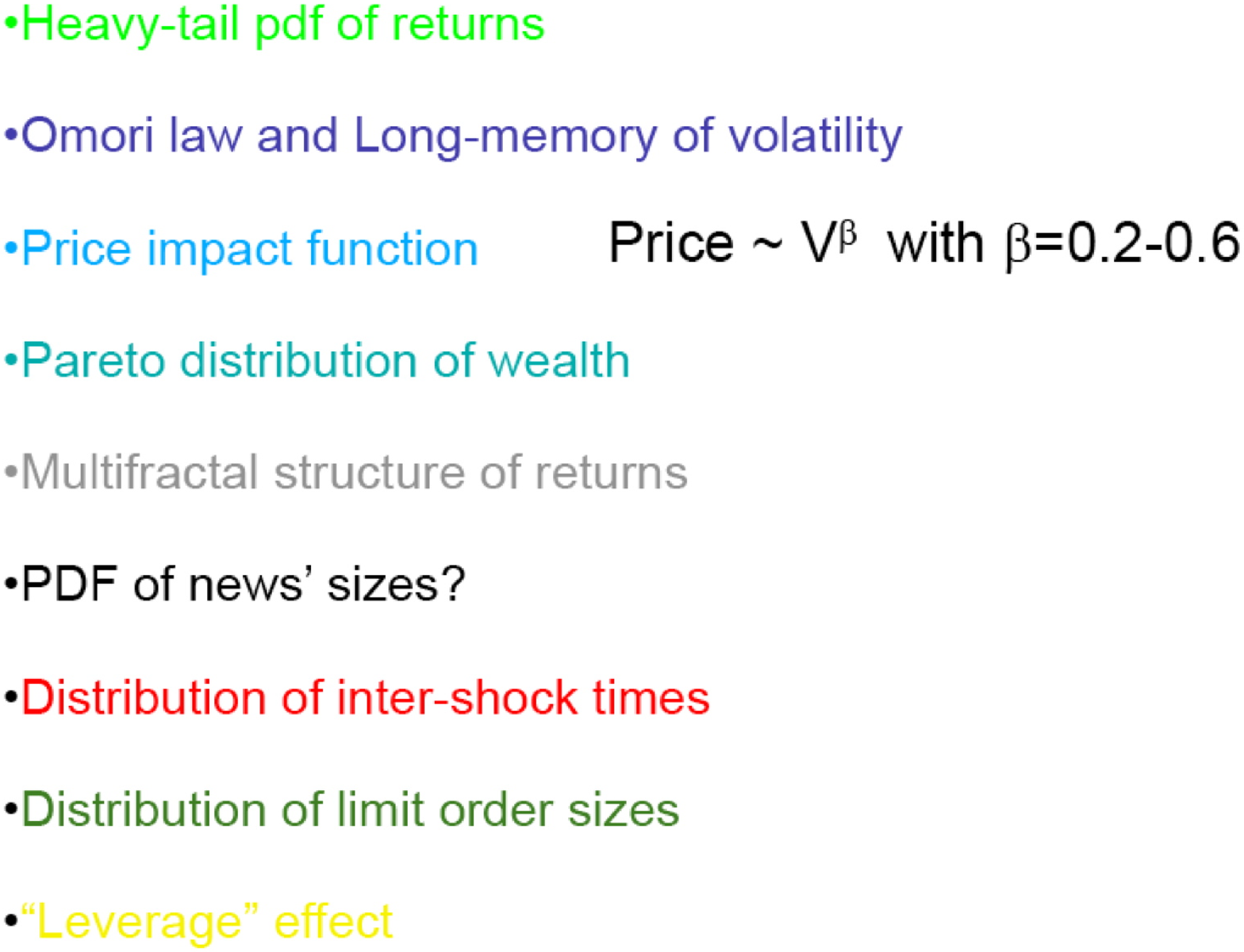}
\caption{Survey of the major statistical laws in financial markets. The color code allows for comparison
with the statistical laws in seismicity and in epileptic seizures documented in figures 
\ref{fig5_statlawEQ} and \ref{fig7_statlawseizures}.}
\label{fig6_statlawfinance}
\end{figure}

Figure \ref{fig6_statlawfinance} presents the most important statistical laws that characterize
the regularities found in financial time series of returns. 
\begin{enumerate} 
\item The distribution of 
financial returns (or relative price variations) is fat-tailed, with 
a tail approximately described by a power law, but the exponent is in the range $2-4$
and thus much larger than for earthquake energies (whose exponent is $\simeq 2/3$). Hence, the 
returns have a well-defined and variance.
\item The relaxation of the level of activity of price
fluctuations (called financial volatility) after a burst is also found to decay approximately
as a power law \cite{Lillo-Mantegna-Omori,VolMRW}, similarly to the Omori law\index{Omori law} of earthquake aftershocks\index{aftershocks}. 
\item The analog of the productivity law of earthquake is the ``price impact function''
which relates the price change to the volume of stocks of a given transaction:
the larger the demand for a stock, the more the price is pushed up.
\item Prices fluctuate because investors place orders. The size of the orders play
in important role, as just said. The sizes of orders are obviously related to the 
``sizes'' of the investors: a large mutual fund managing $100$ billion dollars
has much more impact on the market that an individual managing a modest portfolio.
The size distribution of individuals' wealth, of firm sizes, of mutual fund portfolios,
or university endowments are all found to be power laws. Characterizing
the distributions by the probability density function (pdf), it is found of the form $\simeq 1/W^f$ 
with exponent close to $2$, which corresponds to Zipf's law \cite{SaichevetalZipf09}. For such exponents, 
the mean is either not defined or converge poorly in typical statistical estimations.
\item The size distribution of portfolios plays a role similar to the fault distribution in earthquakes:
portfolio sizes impact the size and nature of orders that move prices; reciprocally, the 
cumulative effect of price moves controls the performance of investment portfolios, and thus
whether the size increases or decreases. We encounter again the chicken-and-egg
structure.  
\item There is also ample evidence that financial time series of returns 
are characterized by multifractal scaling. 
\item The analogy of stress would be news but
we are only starting to understand what is a ``news'' size and how to quantify it via
the response function of social networks (see 
\cite{roehner2004news,deschatres2005amazon,sornette2005origins,cranesorYouTube}).
\item The distribution of time intervals between high levels of volatility has a similar structure
as the inter-earthquake time distribution. 
\item The distribution of limit-order sizes, analogous
to the distribution of seismicity rates, is also a power law \cite{Gopikrishnan00,Gabaixetal06}.
\item However, the so-called ``leverage effect'', in which past losses (large negative returns),
tend to increase future volatility (and not reciprocally) \cite{PerelloLev}, does not seem to have any
counterpart in seismicity.
\end{enumerate}

\begin{figure}
\includegraphics[width=350pt]{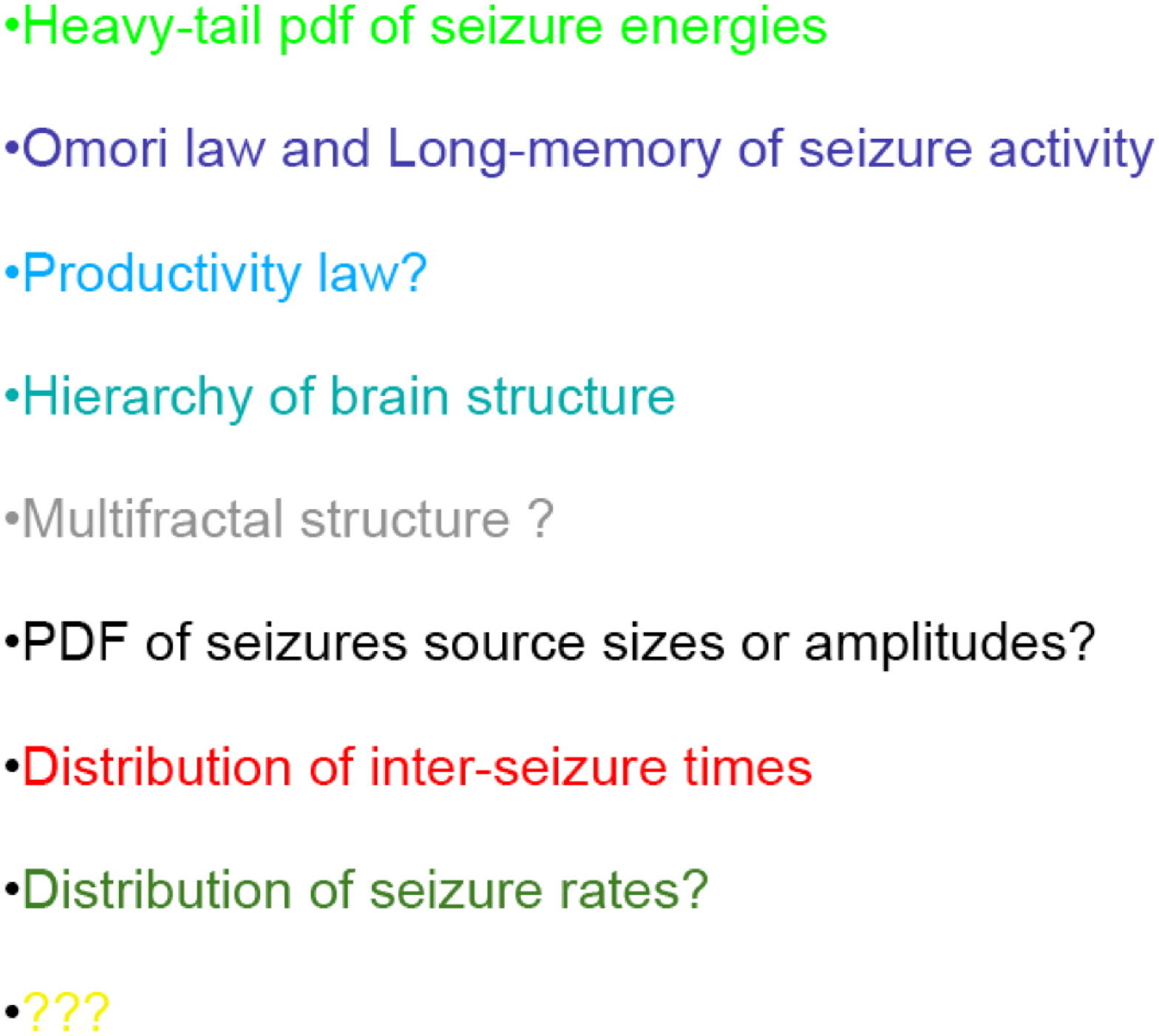}
\caption{Survey of the major statistical laws known in epileptology.  The color code allows for comparison
with the statistical laws in seismicity and in financial markets documented in figures 
\ref{fig5_statlawEQ} and \ref{fig6_statlawfinance}.}
\label{fig7_statlawseizures}
\end{figure}

\begin{figure}
\includegraphics[width=350pt]{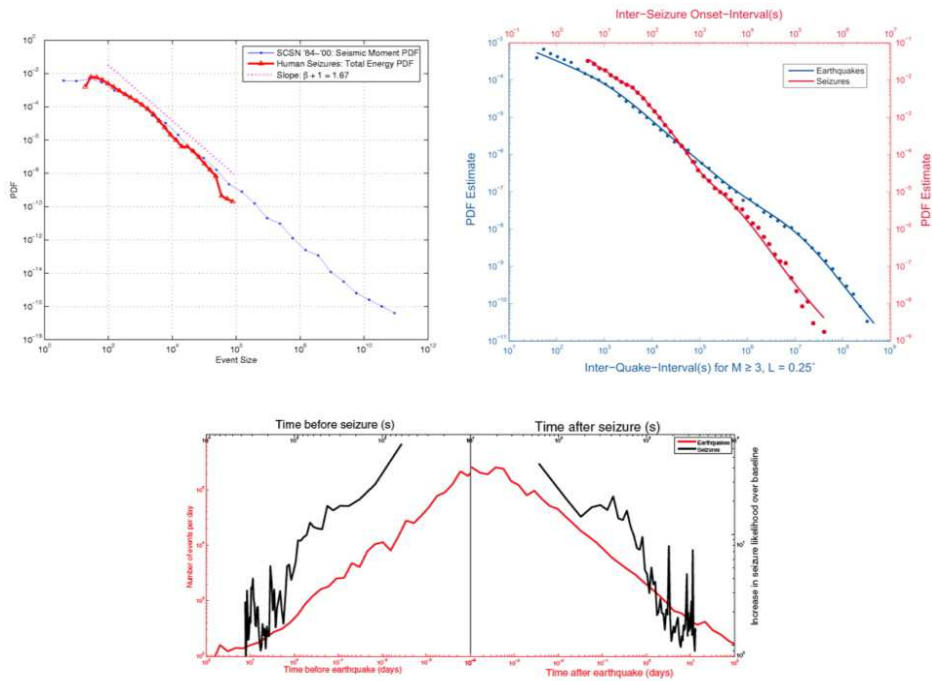}
\caption{Upper left panel: Empirical probability density function (pdf) estimates of seismic moments (SCSN catalog; 1984-2000) (blue curve) and of seizure energies of 60 human subjects (red curve) originating from different epileptogenic regions. 
Upper right panel:  empirical probability density function estimates of the inter-event times between earthquakes (blue circles; blue lower left scales) and seizures in humans (red circles; red upper right scales). 
Lower panel: superimposed epoch analysis of seizures (red line) and earthquakes (blue line) to test for the existence in seizures of ``aftershocks'' (Omori-like behavior) and ``foreshocks''\index{foreshocks} (inverse Omori-like behavior). 
Reproduced from Ref.~\cite{Osorioetal2}.
}
\label{fig8_GR-Omori-pdf_SZ}
\end{figure}

Figure \ref{fig7_statlawseizures} reviews a number of statistical laws 
that have been found to characterize ``focal'' seizures in humans and generalized seizures in animals
\cite{Osorioetal1,Osorioetal2}. 
\begin{enumerate}
\item The analogy with earthquakes is particularly striking for
the Gutenberg-Richter\index{Gutenberg-Richter law} distribution of event sizes, the Omori and inverse Omori laws,
and the distribution of inter-event intervals, as shown in figure \ref{fig8_GR-Omori-pdf_SZ}.
\item While these events occur
in drastically different systems, they 
may nevertheless be described at a coarse-grained level by similar
models of coupled heterogeneous threshold oscillators of relaxation: 
this provides an inspiration to 
investigate the possible existence of other statistical laws, such as productivity. One can suspect
that the triggering\index{triggering} ability of a seizure to promote another future seizures \cite{Osorioetal1,Osorioetal2}
might depend on its duration, amplitude and/or energy.  This remains to be tested.
\item We have already mentioned the hierarchical structure of the brain, as the structure supporting
the spatio-temporal organization of brain activities and the seizures. But it is not known 
whether it can be characterized with multifractal properties. 
\item The analog to stress sources
in earthquakes would be the electric current field within the brain or gaba or other chemical compound 
concentration fields. It remains to be quantified whether these fields present interesting statistical properties,
that may be used to better constrain modeling and perhaps be used for diagnostic.
\item The distribution of seizure rates has neither yet been quantified in a systematic manner. 
\item And there is no obvious analogy with the leverage effect in finance. It is possible that 
for similar asymmetric dynamical effects exist, which would reveal at a collective level the 
asymmetry between excitatory and inhibitory processes in the brain.
\end{enumerate}

\subsection{Rationals for the analogy between earthquakes, financial crashes and epileptic seizures}

The previous section has documented (and also extended conjectures on) a number of quantitative
and qualitative correspondences between earthquakes, financial crashes and seizures.
It is perhaps a priori counter-intuitive to compare earthquakes, financial fluctuations and seizures (the events), 
or fault networks, financial markets and neuron assemblies (the events' supporting structures), 
due to the systems' large differences in scales and in their constituent matter.
However, the proposed correspondence may be motivated and at least partially explained 
on the grounds that these phenomena occur in systems composed of interacting heterogeneous threshold oscillators. 

Consider first the textbook model of an earthquake represents a single fault slowly loaded 
by cm/year tectonic deformations until a threshold is reached at which meter-scale displacements occur in seconds. 
This textbook model ignores the recent realization that earthquakes do not occur in isolation but are part 
of a complex multi-scale organization in which earthquakes occur continuously at 
all spatio-temporal scales according to a highly intermittent, frequent  energy release process 
\cite{kagan94,ouillonSor05}. Indeed, the Earth crust is in continuous jerky motion almost everywhere 
but due to the relative scarcity of recording devices, only the few sufficiently large ones 
are detected, appearing as isolated events. In this sense, the dynamics of earthquakes 
is similar to the persistent barrages of subthreshold oscillations and of action potentials 
in neurons, which sometimes coalesce into seizures. 

Market investors continuously place
limit and market orders, with buyers (respectively sellers) tending to push prices up (respectively down). 
Early on, Takayasu et al. \cite{Takayasu92} noticed that trading strategies lead to dynamics
belonging to the larger class of threshold dynamics with mean-reversal behavior, akin to the outcome
of coupled threshold oscillators of relaxation. Traders and investors enter and exit financial markets
at many different time scales, from milliseconds for the most modern electronic automatic
platforms to years for investors with long horizons. The evolution of their impact is on the order
of years, which is the time scale for growth or decay of fortunes. Furthermore, market rules
and regulations, such as the Glass-Steagall act of 1932-33 or the Sarbanes-Oxley act of 2002, appear
as reactions to extreme market regimes such as financial crashes (the 1929 crash and ensuing
depression for the former and the accounting scandals revealed by the collapse of market
capitalization of new technology firms in 2000), illustrating another process for the evolution
of supporting structures co-evolving to the dynamics of events. 

The separation of time scales in epileptogenic neuronal assemblies is similar
(milliseconds to years) to financial markets (milliseconds to years), but smaller
than in fault networks (fraction of seconds to millenia),
but the organization of coupled threshold oscillators is not very sensitive to the magnitude of the separation 
of time scales, as long as there is one, a property that characterizes relaxational processes. 

The term ``relaxational process'' is here applied to phenomena with a disproportionately 
long (hours to years) charging/loading process vis-a-vis the very short (seconds to minutes) discharge 
of the accumulated seismic energy, money/assets or neuronal  membrane potentials. For instance, in the case of earthquakes, 
the slow motion of tectonic plates at typical velocities of a few cm/year accumulates strains in the core 
of locked faults over hundreds to thousands of years, which are suddenly relaxed 
by the meter-size slips occurring in seconds to minutes that define large earthquakes.  
Thus, one fault taken in isolation is genuinely a single relaxation threshold oscillator, 
alternating long phases of loading and short slip relaxations (the earthquakes). While less well 
studied than earthquakes, the long (hours to years) interval between seizures and their short 
duration (rarely over 2 min) interpreted in light of the fact that the brain is composed of 
relaxational threshold oscillators (neurons) supports the notion that seizures 
too are also relaxational phenomena. The relaxation nature of investment dynamics
can be seen as the result of the competition between different strategies 
available to each investor and their collective output.  This is particularly evident
for first-entry games \cite{Rapoportetal98} and minority games \cite{ChalletMarZhangbook,Coolenbook},
in which agents with bounded rationality are continuously oscillating between different strategies,
creating collectively large market price fluctuations and crashes.

\subsection{Generic phase diagram of coupled threshold oscillators of relaxation}

It is well-known in statistical physics and in dynamical systems theory that ensembles 
of interacting heterogeneous threshold oscillators\index{threshold oscillators} of relaxation generically exhibit self-organized 
behavior with non-Gaussian statistics \cite{rundleetal95,zhaochen02,Kapirisetal05}.
The cumulative evidence presented in figures \ref{fig5_statlawEQ}, \ref{fig6_statlawfinance}
and \ref{fig7_statlawseizures} provides a strong case for the dynamical analogy between earthquakes,
financial fluctuations and seizures, i.e., the existence of an underlying universal organization 
principle captured by the sand pile avalanche paradigm and the concept of self-organized criticality\index{self-organized criticality} \cite{Bak96}. 

\begin{figure}
\includegraphics[width=350pt]{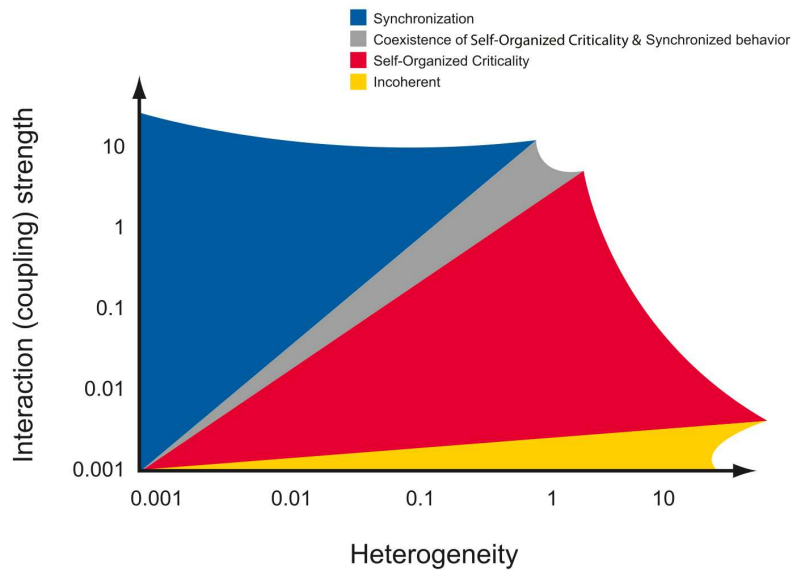}
\caption{Qualitative phase diagram illustrating the effect of changes in coupling strength (y-axis) and heterogeneity (x-axis) on the behavior of systems (such as the brain) composed of interacting threshold oscillators. Marked increases in excitatory coupling drives the system towards the synchronized regime. Slight increases in coupling drive the system towards the power law regime indicative of self-organized criticality. Reproduced from Ref.~\cite{Osorioetal2}, itself adapted from Ref.~\cite{SorMilVan95}.}
\label{fig9_diagram}
\end{figure}

A generic qualitative phase diagram (Figure \ref{fig9_diagram}) depicts the main different regimes 
found in systems made of heterogeneous coupled threshold oscillators, such as sandpile models, 
Burridge-Knopoff block-spring models \cite{Schmittbuhletal93} and earthquake-fault models 
\cite{Soretal94,SorMilVan95,Dahmenetal98}: 
a power law regime (probably self-organized critical) (Figure \ref{fig9_diagram}, right lower half) is 
co-extensive with one of synchronization \cite{strogatz04} 
with characteristic size events (Figure \ref{fig9_diagram}, upper left half). 
This phase diagram embodies the principal qualitative modes that result from the ``competition'' 
between strong coupling leading to coherence and weak coupling manifesting as incoherence.  
Coupling (or interaction strength) is dependent, among others, upon features such as the distance 
between constituent elements (synaptic gap size in the case of neurons), their type  (excitatory or inhibitory) 
and extent of contact (number of synapses and their density), the existence and size of delays in the transmission of signals 
as well as their density and flux rate between constituent elements. 
Heterogeneity, the other determinant of the systemÕs organization, may be 
present in the natural frequencies of the oscillators (when taken in isolation),
in the distribution of the coupling strengths between pairs of oscillators, 
in the composition and structure of the substrate (earth or neuropil) and in their topology among others.  
As shown in Figure \ref{fig9_diagram}  for very weak coupling and large heterogeneity, 
the dynamics are incoherent; increasing the coupling strength (and/or decreasing the heterogeneity) leads to the emergence of intermediate coherence and of a power law regime (self-organized criticality (SOC));
further increases in coupling strength (and/or decreases in heterogeneity), force the system towards strong coherence/synchronization and periodic behavior. 

The specific boundaries between these different regimes depend on the system under study and on the details of the constituting elements and their interactions. In addition, these boundaries may have multiple bifurcations across a hierarchy of partially synchronized regimes within the system. The diagram of Figure \ref{fig9_diagram}
is adapted from the study of a system of coupled fault elements subjected to a 
slow tectonic loading with quenched disorder in the rupture thresholds \cite{SorMilVan95}. In the SOC regime, 
the extreme events are not different from smaller ones, making the former 
practically unpredictable or at most very weakly predictable \cite{Talebswan07}. 
In contrast, in the synchronized regime, the extreme events are different, 
i.e., they are outliers or ``dragon-kings'' \cite{LahSor98,Dragonkings09} occurring as a result of some additional amplifying 
mechanism; these outliers unlike those in the SOC regime, have a degree of predictability
\cite{SorPredictPNAS02}, as we discuss below. 

The model described in section \ref{sectLG} constitutes a nice example of a system that can be described 
by the phase diagram shown in figure \ref{fig9_diagram}. The correspondence
works as follows:
\begin{itemize}
\item The heterogeneity dimension
corresponds to the amplitude of the noise $n$ defined in equation (\ref{thuyj6ujujk}).  
\item The coupling strength 
is quantified by the ratio ${\chi \over \alpha}$ of the instability growth rate divided
by the diffusive relaxation rate.
\end{itemize}
A large ratio  ${\chi \over \alpha}$ corresponds to a large
coupling strength because the local order parameter $S(r, t)$ then exhibits large fluctuations
because the full amplitude between the two branches of the subcritical pitchfork bifurcation
can be sampled, and these large fluctuations have proportionally
a strong influence on neighboring locations. This rationalized the results
that dragon-kings emerge only for relatively small noise levels $n$ and large ratios ${\chi \over \alpha}$.

\section{Self-excited Hawkes process for epileptic seizures}

The analogy with earthquakes and financial fluctuations, and in particular the evidence that seizures may trigger\index{triggering} other 
seizures (inverse and direct Omori laws shown in figure \ref{fig8_GR-Omori-pdf_SZ}), motivates
the presentation of a class of stochastic processes that is specifically formulated to account for triggering,
also called ``self-excitation.'' But, before diving in the formalism, some caveats and definitions must be addressed.

\subsection{``Particles'' versus ``waves''}

\begin{figure}
\includegraphics[width=350pt]{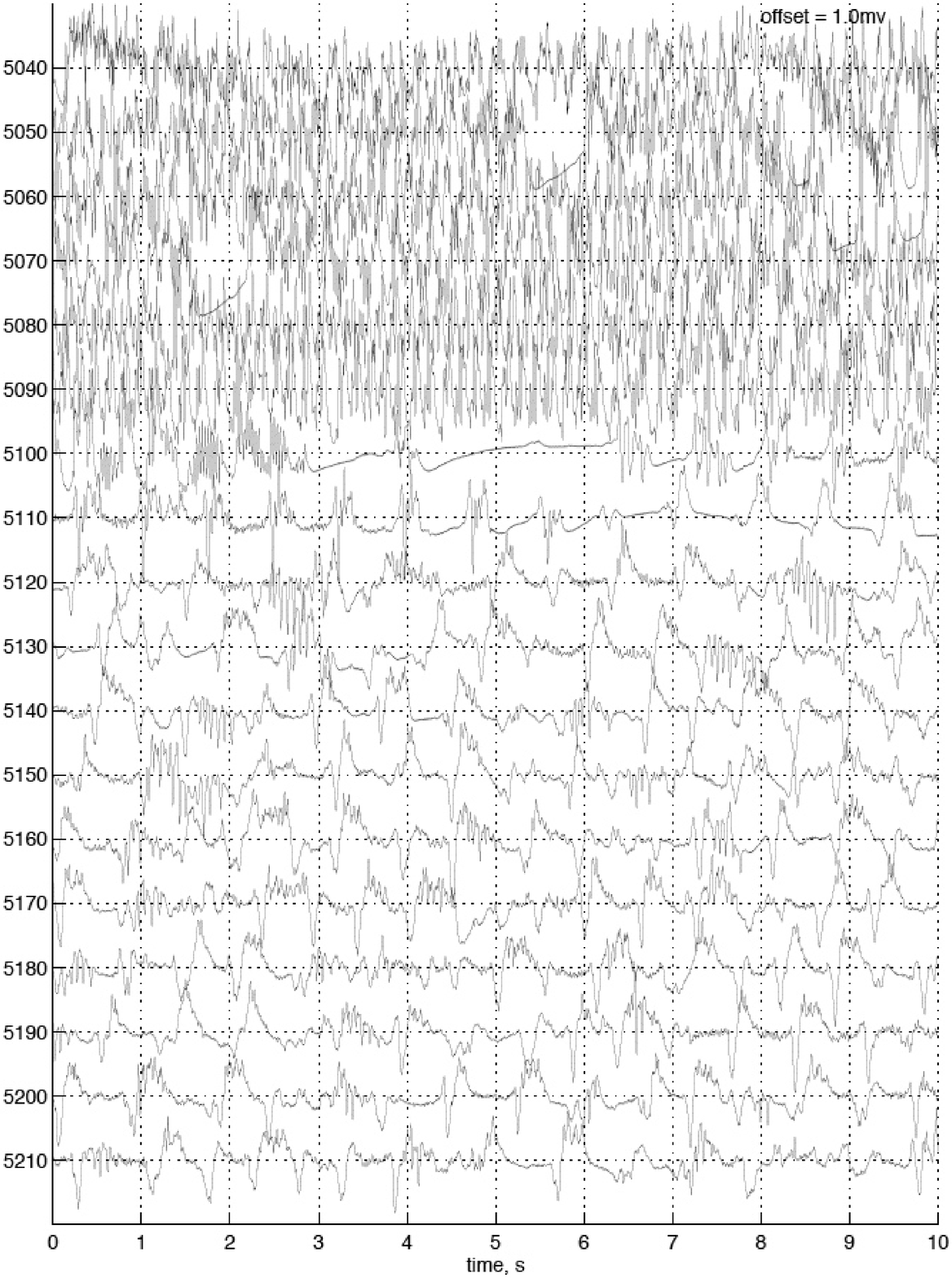}
\caption{Continuous voltage recorded directly from the brain of a rat which received injections
of an epileptogenetic substance.
}
\label{Figseizureburst}
\end{figure}

While clinical seizures are rather unambiguous objects on the basis of the 
often dramatic observable symptoms, 
continuous voltage recordings directly from the brains of human subjects
(electrocorticogram, ECoG) show the existence of many so-called
sub-clinical seizures \cite{OsorioFreiWilk98,Osorioetal02}, i.e., ECoG patterns that are undistinguishable
from their clinical siblings (except perhaps for their
durations and extend of spread) but without obvious manifestations. In textbooks,
``ictal'' events are classified as having clinical manifestations and
interictal events  as lacking visible behavioral changes.
not, in the usual sense of clinical manifestations. But 
the definition and characterization of relevant patterns that can be used
for diagnosing incoming clinical seizures remains elusive. For instance, the 
above textbook concepts of ``ictal'' and ``interictal'' events turn out to be quite fuzzy,
given the demonstration that their durations do not form two well-separated
classes (long durations for ictal events and short duration for interictal events)
but a continuum better characterized by scale-free power law statistics
\cite{Osorioetal1,Osorioetal2}. In addition, so-called interictal events
comprise additionally what have been coined as ``spikes'' and ``bursts of spikes''.  Figure \ref{Figseizureburst}
shows a trace of a continuous recording from the brain of a rat which received injections
of a convulsant. One can observe at the top a pattern
that qualifies as an epileptic seizure, followed by bursts of spikes or by single spikes.
In some cases, interictal spikes
appear to arise from a different location (in a given brain) from the site
of seizure initiation, which has led some to propose that they are quite
distinct mechanistically. As better recording methods are available
and longer time series of ECoG provide data for more sophisticated
statistical analyses, understanding the relationship between spikes, bursts and seizures
is highly relevant, given the growing realization of the 
fuzziness of past classifications based mainly on clinical criteria. Moreover, one should not exclude 
the possibility that spikes and bursts could be relevant diagnostics or even precursory signals
announcing clinical seizures, since they also constitute signatures of the 
excitatory activity of the brain.

In the following, we formulate a model of self-excitation that remains as general
as possible, keeping open the possibility for interactions between spikes, bursts and seizures.
Similarly to earthquakes or financial crashes, the key idea is to view the
activity of a brain, as measured by electrocorticograms, as a wave-like background
on which particle-like structures appear and possibly interact. We refer
to this view as the ``particle'' approach, as opposed to the ``wave'' approach.
The ``wave'' approach consists in viewing the ECoG as a continuous signal and then
apply various signal analysis techniques, for instance derived from 
the theory of dynamical systems and chaos \cite{Lehnertzetal07}.
In contrast, the ``particle'' approach assumes that coherent structures
or patterns exist on the noisy ``wave'' background, allowing to treat
them as individuals or events. The formalism is then constructed to describe
the relationships between these discrete events. 

\subsection{Brief classification of point processes}

When using the ``particle'' point of view, the relevant mathematical language is that of so-called
point processes (also known as shot-noise in physics
or jump processes in finance). Daley and Vere-Jones provide a rigorous development of the theory of point processes
\cite{DaleyVereJones}.

The (conditional) rate $\lambda(t | H_t)$ (also called ``conditional intensity'') of a point process is defined by
\begin{equation}
\lambda(t | H_t) = {\rm lim}_{\Delta \to 0} ~~{1 \over \Delta} {\rm Pr}({\rm event~ occurs~in}~ [t, t+\Delta] | H_t)~,
\end{equation}
where Pr$(X)$ means ``probability that event $X$ occurs.'' The symbol
$H_t$ represents the entire history up to time $t$, which includes all previous events.
This definition is straightforward to generalize for space-dependent intensities
 $\lambda(t,{\vec t} | H_t)$ and to include marks such as amplitudes or magnitudes (see below). 
 The Poisson process is the special case such that $\lambda(t | H_t)$ is constant. Recall that
the simplest point process is the memoryless Poisson process, 
in which events occur continuously and independently of one  another. 
The term ``conditional'' refers to the fact that, in general, the rate $\lambda(t | H_t)$ 
is not constant but may depend on the past history, i.e., on the specific realization of 
past events.

Let us define $f(t | H_t)$ as the probability density function (pdf) time until the next event
(possibly dependent on more than just the last event, when the process in non-Markovian)
and $F(t | H_t)$ as the corresponding survivor function (or complementary
cumulative distribution function). The relationship
between the conditional intensity and these two quantities is given by
\begin{equation}
\lambda(t | H_t) = { f(t |H_t) \over F(t | H_t)}~.
\end{equation}
The probability of an event in the time interval $[t_c, t_c+s]$ is given by
\begin{equation}
P(t_c; s | H_{t_c}) = 1 - \exp \left( -\int_{t_c}^{t_c+s} \lambda(u | H_{t_c}) du \right)~.
\label{hjuyjrym}
\end{equation}
When an event occurs, the history $H_t$ changes and therefore $\lambda(t | H_t)$
may change abruptly, as it is defined as a piecewise continuous function between events.
Another useful relationship relates the pdf $f(t_i | H_t)$ for the $i$-th event to the
conditional density, by differentiation of equation (\ref{hjuyjrym}):
\begin{equation}
f(t_i | H_{t_i}) =  \lambda(t_i | H_{t_i})  \exp \left( -\int_{t_{i-1}}^{t_i} \lambda(u | H_{u}) du \right)  ~ {\cal I}(t_i - t_{i-1})~,
\end{equation}
where ${\cal I}(\cdot)$ is the Heaviside step function. 

\vskip 0.5cm
\noindent
$\bullet$ {\bf Renewal processes}: 
Renewal processes constitute the simplest class of point processes.  
A renewal process is a particular class of 
temporal point process in which the probability of occurrence of the next event
depends only on the time since the last event. The pdf of the waiting time
from the $(i-1)$-th event to the $i$-th one is defined by
\begin{equation}
f(t_i | H_{t_i} = f(t_i | t_{i-1}) = f(t_i-t_{i-1}) ~ {\cal I}(t_i - t_{i-1})~,
\end{equation}
expressing the fact that the history $H_{t_i}$ is reduced to the knowledge of $t_{i-1}$.
Renewal processes are to point processes what are Markov processes
to general stochastic processes.

One can equivalently defined renewal processes by the fact that their conditional intensity at time $t > t_i$,
where $i$ is the index of the last event,
depends only on the occurrence time of the last event $t_i$:
\begin{equation}
 \lambda(t_i | H_{t_i}) =  \lambda(t -t_i)~.
 \end{equation}
 The Poisson process is the simplest renewal process, and corresponds to the
 specification
 \begin{equation}
 f_{\rm Poisson}(\tau; \lambda) = \lambda \exp(-\lambda \tau)~,
 \label{yjujueh}
 \end{equation}
 where we note $\tau=t-t_i$, the running time since the last event $i$.
 This exponential form of the Poisson process is uniquely associated with 
 its memoryless property, which can be quantified by asking for instance
 ``what is the average remaining waiting time at present time $t$, given that a time
 $t-t_i$ has passed since the last event?''  It turns out that the Poisson process
 is the only process such that the average remaining time remains equal to 
 $1/\lambda$ at all times $t$. The conditional distribution of the remaining time,
 conditional on having already waiting the time  $t-t_i$ since the last event
 also remains unchanged in the form (\ref{yjujueh}). 
 Sornette and Knopoff have offered a systematic classification of renewal
 processes into three classes \cite{SorKnopoff97}.
 \begin{enumerate}
 \item When the pdf $f(\tau)$ has a tail decaying faster than exponential, 
 the longer the time since the last event, the shorter the average remaining waiting time
 till the next event.
 \item When the pdf $f(\tau)$ has an exponential tail, the average remaining waiting time
 till the next event is independent of the time that has elapsed since the last event
 (this is the Poisson process).
 \item When the pdf $f(\tau)$ has a tail decaying slower than exponential, 
 the longer the time since the last event, the longer the average remaining waiting time
 till the next event.
 \end{enumerate}
 These statements can be made more precise by calculating explicitly the full shape
 of the conditional distribution of waiting time till the next event, conditional 
 on having already waiting some time $t-t_i$ since the last event that occurred at $t_i$. See 
 Ref.~\cite{SorKnopoff97} for detailed information. Osorio et al. 
 have used this statistics as one of the diagnostics to characterize the sequence
 of epileptic seizures and to compare with earthquake sequences \cite{Osorioetal1,Osorioetal2}.
 
\vskip 0.5cm
\noindent
$\bullet$ {\bf Clustering models}: These models capture the general observation
for earthquakes, financial volatility and seizures that they occur in bursts, 
that is, according to patterns exhibiting much more clustering or grouping than
predicted by renewal processes.

Clustering models are usually constructed from two processes: a cluster center process, 
which is often a renewal process, and a cluster member process. In simple terms, 
the center events are main events or sources, from which the member events derive.
The cluster member process consists of events that are triggered\index{triggering} by the cluster centers via a triggering
function $h(t-t_i, \xi)$, which usually depends only on the time $t-t_i$ since the occurrence
time $t_i$ of the cluster center, and on a stochastic amplitude $\chi$ drawn from a distribution
usually chosen to be invariant in time. In other words, cluster centers are parents
and the cluster members are their corresponding offsprings: a given parent triggers
only his cohort of offsprings and has no influence on the offsprings of other parents (center sources). 

An example is given by the simple aftershock\index{aftershocks} model,
which considers that there are main shocks distinctly different from their aftershocks.
The former are the cause of the later, which cluster strongly after them. Such aftershock
model is the standard textbook model for main earthquakes and their aftershocks. It consists
in writing the conditional intensity as
\begin{equation}
\lambda(t | H_t^c, \Theta) = \lambda_c + \sum_{i_c | t_{i_c} < t}  h(t-t_i, \xi)~,
\label{hyjuj}
\end{equation}
where $H_t^c$ is the history up to time $t$ that needs only include information
about the cluster centers $\{ t_{i_c}, \chi_{i_c}\}_{1 \leq i_c \leq N}$, as cluster
members do not trigger their own events and do not influence the future.
In the specification (\ref{hyjuj}), we have assumed for simplicity that the cluster center process
is a Poisson process with constant rate $\lambda_c$. The triggering process
from centers to members is described by the set of parameters $\Theta$ characterizing
the kernel $h(t-t_i, \xi)$ quantifying the ability of centers to trigger their offsprings. 

\vskip 0.5cm
\noindent
$\bullet$ {\bf Self-excited\index{self-excited point process} models}: These models were first introduced by Hawkes\index{Hawkes process} \cite{Hawkes1971a,Hawkes1971b}
and Hawkes and Oakes \cite{HawkesOakes1974}. They generalize the cluster models by allowing each event, including
cluster members, i.e., aftershocks, to trigger their own events according to some memory
kernel $h(t-t_i, \xi)$.
\begin{equation}
\lambda(t | H_t, \Theta) = \lambda_c + \sum_{i | t_{i} < t}  h(t-t_i, \xi)~,
\label{hyjuetg2tgj}
\end{equation}
where the history $H_t = \{ t_i \}_{1 \leq i \leq N}$ now includes all events and the sum
in expression (\ref{hyjuetg2tgj}) runs over all triggered events. The term $\lambda_c$ means
that there are still external background sources occurring according to a Poisson process
with constant intensity $\lambda_c$ but all other events can be both triggered
by previous events and can themselves trigger their offsprings.  This gives rise
to the existence of many generations of events.

\vskip 0.5cm
\noindent
$\bullet$ {\bf Marked self-excited point processes}: This class is a multidimensional
extension of the former self-excited process. The generalization consists in associating
with each event some marks (possible multiple traits), drawn from some distribution $p(m)$,
usually chosen invariant as a function of time:
\begin{equation}
\lambda(t, m| H_t, \Theta) = p(m) \left( \lambda_c + \sum_{i | t_{i} < t}  h(t-t_i, \xi, m_i) \right)~,
\label{hyjuetgq2h56h2tgj}
\end{equation}
where the mark $m_i$ of a given previous event now controls the shape and properties
of the triggering\index{triggering} kernel describing the future offsprings of that event $i$. 
The history now consists in the set of occurrence times of each triggered event and their 
marks: $H_t = \{ t_i, m_i\}_{1 \leq i \leq N}$. The first factor $p(m)$ in the r.h.s. of 
expression (\ref{hyjuetgq2h56h2tgj}) writes that the marks of triggered events 
are drawn from the distribution $p(m)$, independently of their generation and waiting times.
This is a simplifying specification, which can be relaxed. Inclusion of spatial kernel
to describe how distance impacts triggering efficiency is straightforward.

\begin{figure}
\includegraphics[width=350pt]{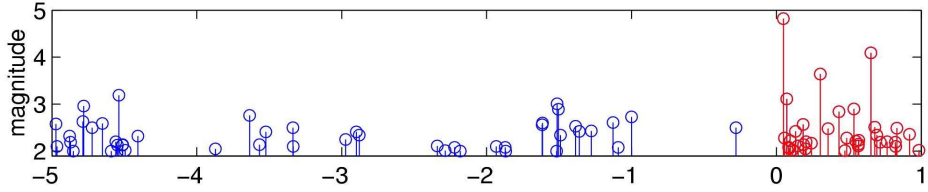}
\caption{Typical realization of a sequence of events using the ETAS model. The horizontal axis is time
and the vertical axis shows the magnitude $m$ (or mark) of each event.}
\label{FigETASgeneration}
\end{figure}

A particularly well-studied specification of this class of marked self-excited point process
is the so-called Epidemic-Type-Aftershock-Sequence (ETAS) model \cite{KK81,Ogata88}:
\begin{equation}
\lambda(t, m | H_t, \Theta) = p(m) \left( \lambda_c + \sum_{i | t_{i} < t}   { k e^{a(m_i-m_0)} \over (t-t_i + c)^{1+\theta} }  \right)~,
\label{yj4ujbg}
\end{equation}
where $p(m)$ is given by the Gutenberg-Richter\index{Gutenberg-Richter law} law with exponent $\beta$ discussed already in section \ref{tyjrukjiuk}.
The memory kernel is chosen as the power law (called the Omori law)\index{Omori law} with exponent $1+\theta$. The lower magnitude
cut-off $m_0$ is such that events with marks smaller than $m_0$ do not generate offsprings.
This is necessary to make the theory convergent and well-defined, otherwise the crowd of small events may actually
dominate. The time constant $c$ ensures normalization and finiteness of the triggering rate immediately 
following any event. Each event (of magnitude $m$) triggers 
other events with a rate $\sim e^{a m}$, which defines the so-called fertility or productivity law.
 The set of parameters is $\Theta = \{ \beta, \lambda_c, k, a, m_0, c, \theta \}$. 
 Figure \ref{FigETASgeneration} shows a typical realization of a sequence of events 
 generated with the ETAS model.
 
An observed ``aftershock''\index{aftershocks}
sequence in the ETAS model is the sum of a cascade of events in which
each event can trigger more events. The triggering process
may be caused by various mechanisms
that either compete with each other or combine.
The ETAS model is parsimonious as it lumps all the complications of 
physical and biological properties as well as  geometric structural geometry in a few key parameters quantifying 
the Omori law, the Gutenberg-Richter law and the productivity law.
This is particularly important as seismic as well as seizure data
is relatively sparse, has  limited precision accuracy, and the characterization
of the properties of these dynamical processes is bound to be full of misleading paths if
solid theoretical and analytical guidelines do not constrain the research
on empirical data.

This class of marked self-excited point processes is now considered as the benchmark
that best describes the statistical properties of spatio-temporal earthquake catalogs.
In particular, the textbook classification of foreshocks\index{foreshocks}, mainshocks and
aftershocks is now considered obsolete by many seismologists, due to
the cumulative evidence that any earthquake may trigger
other earthquakes through a variety of physical mechanisms but this does
not allow one to put a tag on them \cite{FelzerLanders02}. 
The textbook classification of foreshocks, mainshocks and
aftershocks is essentially a human-made construction that is open to revision
as a function of the development of the sequences of earthquake magnitudes.
For instance, if an aftershock happens to have a larger magnitude that
the earthquake that was qualified previously as a mainshock, it is reclassified as
the new mainshock of the unfolding sequence and the previous mainshock
becomes one of its foreshock. The fact that many small earthquakes occur
after large mainshocks, and are thus classified as aftershocks,
is simply due to the fact that large earthquakes trigger many earthquakes and
most earthquakes are small. Thus, it is improbable (but not impossible) that a large earthquake
is followed in close succession by still larger earthquakes.  

Rather than keeping the
textbook classification that foreshocks are precursors of mainshocks and
mainshocks trigger aftershocks, the self-excited class of models starts from the
hypothesis that a parsimonious description of seismicity does not
require the division between foreshocks, mainshocks and aftershocks that
are indistinguishable from the point of view of their physical
processes \cite{FelzerLanders02,Felzer3} (it is however sometimes convenient to use
the time-honored foreshocks-mainshocks-aftershocks terminology, as long as 
it is understood that the model
refers only to events which may trigger other earthquakes). But the story is not
written as, recently, some evidence of a difference between spontaneous and triggered
earthquakes was obtained  \cite{ZhuangChristophersenetal08}.

We propose that a similar approach may be a useful starting point
in epileptology. Single epileptiform discharges (spikes), bursts of spikes and seizures 
may not be, as often claimed, distinct phenomena but simply reflect the heterogeneous 
manifestations of processes governed by the same mechanisms or laws 
while having a self-triggering capacity or degree of ``fertility.'' This seemingly radical 
shift in conceptualization may provide a deeper and more fruitful 
insight into the dynamics of ictiogenesis.

The ETAS model\index{ETAS model: epidemic-type aftershock sequence model} and other related models developed
on similar principles are popular with statisticians interested
in the characterizations of complex spatio-temporal patterns (in particular
with applications to seismicity)
\cite{Kagan91,Musvere92,Rathbun,Ogata98,console,Zhuang1,console2,
Ogakat03,Ogata04,Zhuang2}, using maximum likelihood methods for
parameter estimations and residual analysis \cite{Ogata04,Ogakat03} for the
detection of deviations from normal seismicity. We believe that these
statistical techniques could be usefully applied to seizure time series.

A detailed understanding of the observable properties of the 
marked self-excited point processes has been developed in the last decade, that we briefly summarize below.

\subsection{Main properties of the ETAS model}

We stress that the advantage of the ETAS 
model is to offer a very parsimonious description
of the complex spatio-temporal organization of systems characterized
by self-excitation of ``bursty'' events, 
without the need to invoke ingredients other than the well-documented
stylized facts reported above: distribution of event sizes, Omori law and productivity law. 
An important insight is that the Omori law may come in different forms, which 
can be derived from the same model only via the change of a crucial parameter, 
the branching ratio $n$. The parameter $n$ is defined as the mean number of events of first 
generation triggered per event. Using the notation of expression (\ref{yj4ujbg}), the
branching ratio is given by 
\begin{equation}
n = K {\beta \over \beta -a}~,~~~~~~{\rm where}~ K := {k \over \theta c^\theta}~.
\label{eqpourn}
\end{equation}
The variability of the apparent Omori's exponent $p$ is then obtained
as a result of the relative importance 
of cascades of aftershocks, of aftershocks of aftershocks, and so on over
possibly many generations \cite{MarsanLengline08}. The branching ratio $n$ can vary with time
and from location to location. In the context of epileptic seizures, it can be 
used as a diagnostic of the susceptibility of the brain to trigger epileptic seizures.

While the results summarized below pertain to earthquakes, the method used to obtain them
can be applied to seizure time series as well as financial fluctuations
(for an early attempt in  this later domain, see Ref.~\cite{Chavez-etas05}).

\subsubsection{Subcritical, critical and supercritical regimes}

Precise analytical results and numerical simulations show the existence of 
three time-dependent regimes, depending on the ``branching ratio'' $n$ and on the sign of $\theta$.
This classification
is valid for the range of parameters $a < \beta$. 
When the productivity exponent $a$ is larger than the exponent $\beta$ of the Gutenberg-Richter law,
an explosive regime occurs leading to stochastic finite-time singularities \cite{SorHelmPRL02},
a regime that we do not consider further below, but which is relevant
to describe the accelerated damage processes leading to global systemic failures
in possibly many different types of systems \cite{SorPredictPNAS02}. 
\begin{enumerate}
\item For $n<1$ (sub-critical regime), the rate of events triggered by a given shock
decays according to an effective Omori power law, characterized by
a crossover from an Omori exponent $p=1-\theta$ for $t<t^*$ to a larger exponent
$p=1+\theta$ for $t>t^*$ \cite{SorSor,helm1}, where $t^*$
is a characteristic time $t^* \sim c / (1-n)^{1/\theta}$, which is controlled by the distance from $n$ to $1$.

\item For $n>1$ and $\theta>0$ (super-critical regime), one finds a transition from an
Omori decay law with exponent $p=1-\theta$ at early times since the mainshock
 to an explosive exponential increase of the seismicity rate \cite{SorSor,helm1,SaichSordissect10}.

\item In the case $\theta<0$, there is a transition from an Omori law with exponent
$1-|\theta|$ similar to the local law, to an exponential increase at large times, 
with a crossover time $\tau$ different from the characteristic 
time $t^*$ found in the case $\theta>0$.
\end{enumerate}

These results may open the road for the discovery of new types of seizure precursors.  
These could include (i) variable $p$-values, in particular the suggestion that a small 
$p$-value may be a precursor of a large event, (ii) relative seizure quiescence in some 
spatial domain preceding the occurrence of large seizures, (iii) exponential increase 
in seizure activity in some other spatial domains preceding large events.

\subsubsection{Importance of small events for triggering other events of any size \cite{Helmstetter03,HKJ1}}

In the context of earthquakes for which the productivity exponent $a$ is estimated
smaller than, but close to, the Gutenberg-Richter exponent $\beta$, 
small events have been found to provide a dominating contribution 
to the overall activity, as their number more than compensates their 
relatively smaller individual impact. This is due to the structure of the model in which all events
can trigger other events. This realization comes as a big surprise
to experts, who have been accustomed to the concept that only large 
and great earthquakes needed to be considered since they overwhelmingly
dominate the overall release of energy in the Earth crust. But not so, for the triggering ability,
as is now understood.  Can there be a similar situation for epileptic seizures,
for whom the myriad single spikes, bursts of spikes and subclinical seizures  
play an important role in the triggering of clinical seizures?

\subsubsection{Effects of undetected seismicity: constraints on the size of the smallest
triggering event from the ETAS model \cite{SW1}}

The mechanism of event triggering together with simple
assumptions of self-similarity, as captured in the simple ETAS
specification, make obligatory the existence of a minimum magnitude $m_0$ below
which events do not or only weakly trigger other events.
It turns out to be possible to estimate an order of magnitude of $m_0$
by noting that the magnitude $m_d$ of completeness of empirical catalogs has
no reason to be the same as $m_0$, and by using diverse empirical data
based on maximum likelihood inversions of observed aftershock
sequences of real catalogs with the ETAS model.
The obtained constraint $m_0 \simeq -1 \pm 2$ is loose and reflects
the many uncertainties in the model calibrations and model errors.

\subsubsection{Apparent earthquake sources and clustering biased by
undetected seismicity \cite{SW2,SW2Saichev}}

In models of triggered-seismicity,
the detection threshold $m_d$ is commonly equated to the magnitude $m_0$ of the
smallest triggering earthquake. This unjustified
assumption neglects the possibility that shocks below the detection
threshold may trigger observable events. 
Distinguishing between the detection threshold $m_d$ and the minimum
triggering earthquake $m_0 \leq m_d$, and considering the branching
structure of one complete cascade of triggered events,
an apparent branching ratio $n_a$
and an apparent background source $S_a$ can be determined
from the exact calculation of the sequence of observed triggered events
with marks above the detection threshold $m_d$. The 
presence of smaller undetected events that are capable of triggering larger 
events is the cause for the renormalization. One could imagine
that triggering between seizures could be similarly renormalized
when not taking into account of structures such as spikes and bursts of spikes
if the later have some triggering effects on seizures.

\subsubsection{Cascades of triggered events}

By comparison between synthetic catalogs
generated with the ETAS model and real seismicity, 
it is now understood that a surprisingly large fraction of earthquakes 
in real seismicity are probably triggered
by previous events. Recent conservative lower bounds suggest
that at least $60\%$, and perhaps up to $99\%$ of earthquakes are triggered
by previous earthquakes \cite{HSimp03,SW1,SW2,MarsanLengline08}.
This fraction is nothing but the so-called average branching ratio $n$
or mean number of triggered event per earthquake, averaged over all magnitudes \cite{HSimp03}. 
In addition, within the picture that earthquakes can trigger
events which themselves trigger new events and so on according to the same
basic physics, then, most triggered events within a sequence
should be triggered indirectly through cascades \cite{HSimp03}. Therefore,
previous observations that a significant 
fraction of earthquakes are triggered earthquakes imply that 
most aftershocks are indirectly triggered by the mainshocks. In the class
of ETAS models, this has the implication that the observed Omori law
is obtained from a renormalization of 
the direct Omori law (describing the direct interactions between triggering
and triggered earthquakes) to the global law with different exponent $p$ \cite{SorSor,helm1}.  
The cascades of secondary triggering provides a mechanism for slow aftershock sub-diffusion
 \cite{helmspa,helmspaouillon} and slow foreshock migration \cite{Helmfore,HSforeexpl03}.

\subsubsection{Other results available for marked self-excited point processes}

A number of other interesting mathematical and statistical results have been derived
for the ETAS model, which show that the model has non standard properties
resulting from the interplay between the triggered cascades and the two
power laws characterizing the distribution of sizes and the productivity process.
These results have been obtained by rigorous mathematical derivations using 
probability generating functions:
\begin{itemize}
\item non-mean field anomalous exponents for the distribution of ``cluster'' sizes
due to the interplay between cascades of generation and the power laws
of productivity and of marks (magnitudes) \cite{SaichevHelmSor05};
\item non-mean field distributions of lifetimes and total number of generations
before extinctions of aftershock sequences
emanating from isolated main shocks \cite{SaichevSorlife04};
\item the distribution of waiting times between events in a given region is
characterized by an approximate power law \cite{SaichevSorPRL04,SaiSor07,SorUtkinSai08};
\item stochastic reconstruction of the genealogy of the cascades of triggered events
\cite{Zhuang1,Zhuang2,MarsanLengline08,SornetteUtkin09}.
\end{itemize}

\subsection{Forecasts using self-excited marked point processes}

The understanding of the importance of cascades of triggered
seismicity has led to important improvements of existing methods of
earthquake forecasts \cite{KaganJack00}, based on variations of the ETAS model, by
taking into account the cascades of secondary triggering
\cite{HSpredic03,HelmKaganJackson05,WernerHelmJackKag09}. 

As a quantitative theoretical check, the
number $r$ of earthquakes in finite space-time windows is often taken as
the target for forecasts: for instance within the RELM (Regional
Earthquake Likelihood Models: www.relm.org) project in Southern
California, a forecast is expressed as a vector of earthquake rates
specified for each multi-dimensional bin \cite{Relm}, where a bin is
defined by an interval of location, time, magnitude and focal mechanism
and the resolution of a model corresponds to the bin sizes. 
The full theory of this observable within the ETAS model has been developed 
using the formalism of generating probability functions (GPF) 
describing the space-time organization of earthquake sequences
\cite{SaiSor06,SSdistkwprl}. The calibration of the theory to the
empirical observations for the California catalog shows that it is
essential to augment the ETAS model by taking account of the
pre-existing frozen heterogeneity of spontaneous earthquake sources.
This seems natural in view of the complex multi-scale nature of fault
networks, on which earthquakes nucleate. The extended theory is able to
account for the empirical observation satisfactorily. In particular, 
the probability density functions $P_{\rm data}(r)$ of the
number $r$ of earthquakes in finite space-time windows for the
California catalog, over fixed spatial boxes $5 \times 5$ km$^2$, $20
\times 20$ km$^2$ and $50 \times 50$ km$^2$ and time intervals $dt
=1,~10,~100$ and $1000$ days have been determined. 
One finds a stable power law tail compatible
with $P_{\rm data}(r) \sim 1/r^{1+(b/\alpha)}$
\cite{SaiSor06,SSdistkwprl}. This result recovers previous
estimates with different statistical methods and for large space and
time windows \cite{Corral,others1,others2}, while proposing a simple and
generic explanation in terms of cascades of triggering of earthquakes.
This example and others \cite{HScom04} show the power of the simple
concept of triggered seismicity to account for many (most?) empirical
observations.

The Working Group on Regional Earthquake Likelihood
Models (RELM) has invited long-term (5-year) forecasts for California 
in a specific format to facilitate comparative testing 
\cite{Field2007a,Field2007b,Schorlemmeretal2007,SchorlemmerGerstenberger2007,Schorlemmeretal2009}. Building on RELM's success,  the Collaboratory for the Study of Earthquake
Predictability (CSEP, www.cseptesting.org) inherited and expanded RELM's mission to regionally and
globally test prospective forecasts 
\cite{Schorlemmeretal2009,Zecharetal2009,Werneretalretro10}.
Many of the competing models are based on concept of earthquake triggering
embodied in the marked self-excited conditional point processes.

New developments for point processes include the adaptation
of data assimilation\index{data assimilation} methods \cite{WernerIdeSor}. Recall that,
in meteorology, engineering and computer sciences, data assimilation is
routinely employed as the optimal way to combine noisy observations with
prior model information for obtaining better estimates of a state, and thus
better forecasts, than can be achieved by ignoring data
uncertainties. Earthquake forecasting as well as seizure prediction, too, suffer from  measurement
errors and from model information that is limited, and may thus gain significantly
from data assimilation. Werner et al. have presented 
perhaps the first fully implementable
data assimilation method for forecasts generated by a
point-process model \cite{WernerIdeSor}.  The method has been tested on a synthetic and
pedagogical example of a renewal process observed in noise, which is
relevant to the seismic gap hypothesis, models of characteristic
earthquakes and to recurrence statistics of large quakes inferred from
paleoseismic data records. In order to address the non-Gaussian statistics of
earthquakes, it was necessary to use sequential Monte Carlo methods, 
which provide a set of flexible simulation-based methods for recursively 
estimating arbitrary posterior distributions. Extensive
numerical simulations have demonstrated the feasibility and benefits of
forecasting earthquakes based on data assimilation. 
The forecasts based on the Optimal Sampling Importance
Resampling (OSIR) particle filter are found significantly better than those of
a benchmark forecast that ignores uncertainties in the observed event
times. We predict that data assimilation will also become an important
tool for seizure predictions in the future.

\subsection{A preliminary attempt to generate synthetic ECoG with the ETAS model}

The following is a modest example of how to generate synthetic time series that look
like electrocorticogram (ECoG) using the ETAS model defined by the conditional
intensity given by expression (\ref{yj4ujbg}). We imagine that the elementary events
are ``spikes'' and that spikes can excite other spikes following the ETAS specification. Sequences of closely occurring
spikes may then define bursts and seizures can perhaps be observed when bursts are sufficiently clustered.

The synthetic ECoG are generated as follows. For a given choice of the parameter set
$\Theta = \{\lambda_c, \beta, n, a, m_0, c, \theta \}$, we generate a time series 
of events $\{ t_i, m_i \}$, in which each event $i$ is characterized by its occurrence time $t_i$
and its mark $m_i$. Note that we use $n$ instead of $k$, but the two are related directly
through expression (\ref{eqpourn}).

Then, we assume that each event $i$ is associated with a ``spike pattern'' in 
a virtual ECoG recording given by
\begin{equation}
F(t-t_i) = {\rm sign}_i  \cdot f_i  \cdot {1 \over \sqrt{2 \pi} \tau_i^3} (t-t_i)\exp \left( - {(t-t_i)^2 \over 2 \tau_i^2} \right)~,
\label{tjukik}
\end{equation}
where
\begin{equation}
f_i = f_0 \cdot 10^{\alpha m_i}~, ~~~~~~~ \tau_i = \tau_0 \cdot 10^{d m_i}~.
\label{thythtbqebvq}
\end{equation}
The signal $F(t-t_i)$ is thus a derivative of a Gaussian function and shows a typical dipole structure
with a positive arch followed by a negative arch or vice-versa, depending on the sign term 
`${\rm sign}_i$'  that is chosen here at random and independently for each event. 
The mark $m_i$ of event $i$ is assumed to control the amplitude $f_i$ of the spike and its
duration $ \tau_i$ according to the expressions (\ref{thythtbqebvq}).

\begin{figure}
\includegraphics[width=350pt]{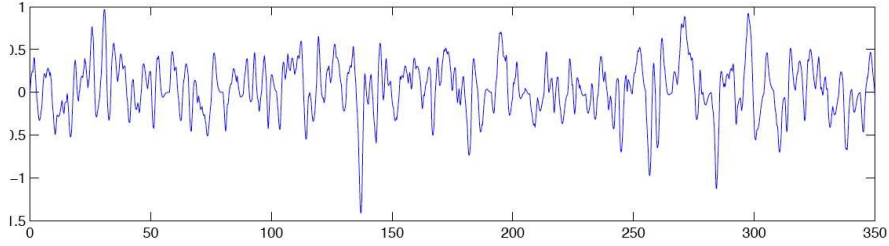}
\caption{Synthetic electrocorticogram constructed with the ETAS model and using the spike pattern
given by expression (\ref{tjukik}) with (\ref{thythtbqebvq}) for the following parameters:
$ \{  \lambda_c=1, \beta=2/3, n=0.5, a=0.2, m_0, c=60, \theta=0.5, a=0.001, d=0.001 \}$.}
\label{FigSynthECoG-ETAS1}
\end{figure}

\begin{figure}
\includegraphics[width=350pt]{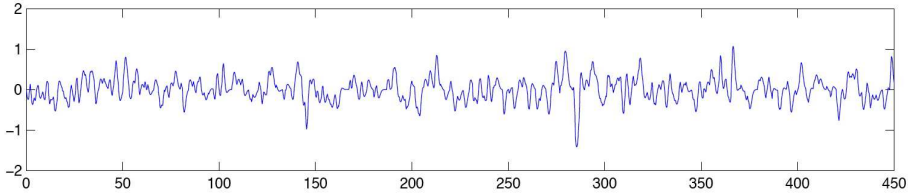}
\caption{Synthetic electrocorticogram constructed with the ETAS model and using the spike pattern
given by expression (\ref{tjukik}) with (\ref{thythtbqebvq}) for the following parameters:
$ \{  \lambda_c=1, \beta=2/3, n=0.5, a=0.2, m_0, c=60, \theta=0.05, a=0.001, d=0.001 \}$.}
\label{FigSynthECoG-ETAS2}
\end{figure}

Figures \ref{FigSynthECoG-ETAS1} and \ref{FigSynthECoG-ETAS2} show two realizations
with the same parameters, except for the memory exponent $\theta=0.5$ in the former
and $\theta=0.05$ in the later. The comparison between the two figures illustrates
the impact of the memory in the triggering of spikes by previous spikes. 
Figure \ref{FigSynthECoG-ETAS1} corresponds to a shorter lived memory and a more 
spiky regime, compared with figure \ref{FigSynthECoG-ETAS2}.

\begin{figure}
\includegraphics[width=350pt]{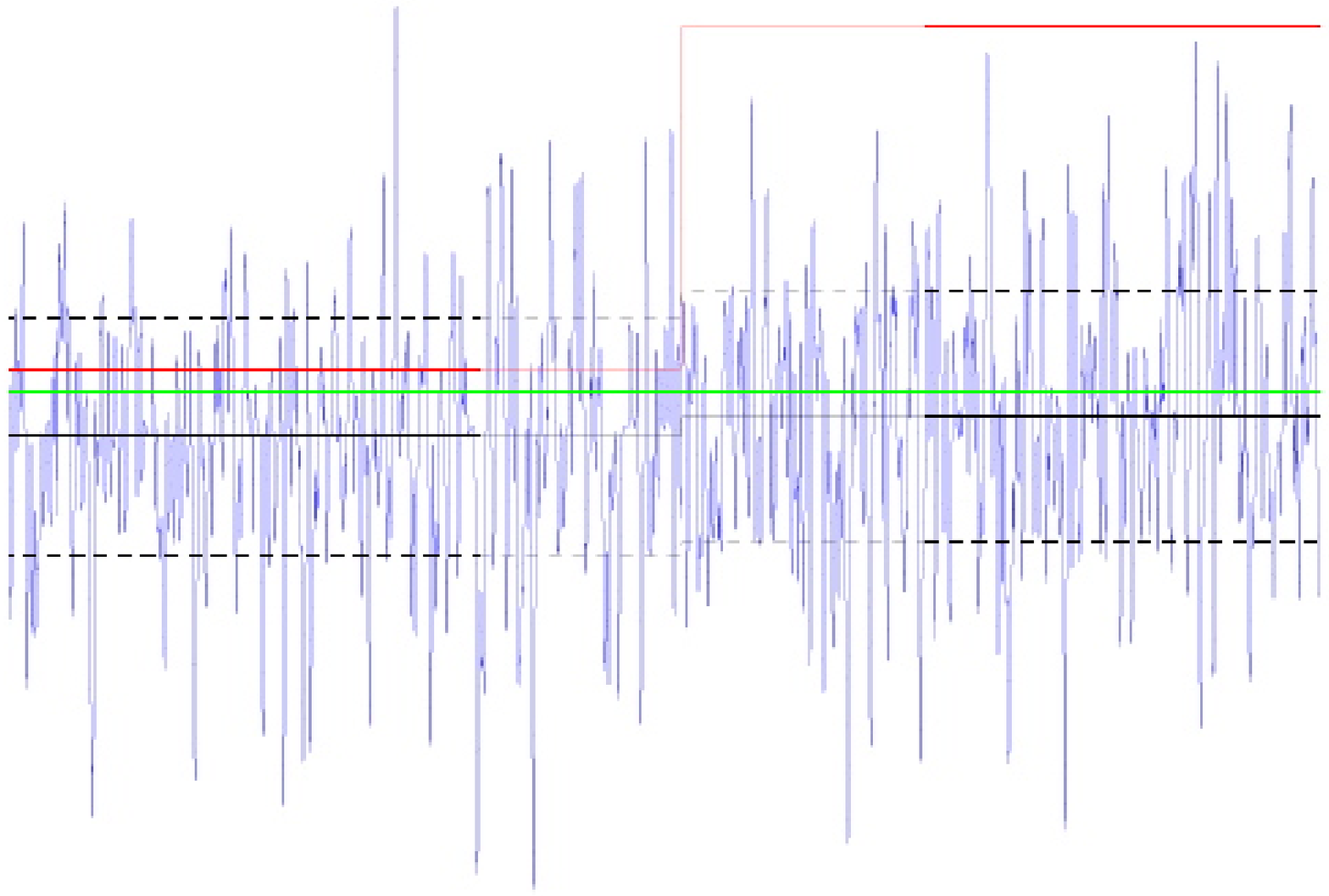}
\caption{Synthetic electrocorticogram constructed with the ETAS model and using the spike pattern
given by expression (\ref{tjukik}) with (\ref{thythtbqebvq}) for the following parameters:
$ \{  \lambda_c=1, \beta=2/3, n=0.1 \to 0.9, a=0.2, m_0, c=60, \theta=0.05, a=0.001, d=0.001 \}$.}
\label{FigSynthECoG-ETAS3}
\end{figure}

Figure \ref{FigSynthECoG-ETAS3} shows a synthetic  electrocorticogram obtained
by changing the branching ratio from a low value $n=0.1$ to a large value $n=0.9$
abruptly in the middle of the graph. For $n=0.1$, the cumulative
effect of ten spikes is needed on average to directly trigger an additional spike. For $n=0.9$, each spike
triggers directly an additional spike almost by itself. This corresponds to a much more intense
activity, with more correlations and burstiness. Seizure-like patterns can be obtained
by decreasing $\lambda_c$ and increasing further the value of $n$ towards
the critical value $1$.

\section{Concluding remarks}

The inherent value of predicting seizures has led to many efforts to fulfill this
aim, efforts that to date have been unsuccessful. In particular, from our point
of view, an acute limitation is the absence of understanding of the spatio-
temporal behavior of seizures and the absence of corresponding models. 
While the reasons behind this state of affairs are multiple, there has
been progress, if only in recognizing the challenges (after a
first wave of over-optimism) as well as the need for rigorous testing 
procedures and for new approaches.

The task of predicting the occurrence of recurrent aperiodic events such
as seizures would benefit from development of models that recognize the 
value of  multi-scale approaches , aimed at excluding features that 
unnecessarily increase algorithmic complexity and the  danger of 
computational irreducibility and its associated unpredictability. This chapterÕs 
central predicates are that seizures may be statistically predictable and that 
application of tools from statistical physics such as renormalization group
 theory may facilitate identification of the scale of observation (likely to be 
coarse-grained) that is the most informative.
Through systematic quantitative comparisons with earthquakes, 
conceptual groundwork (neurons and faults are treated as coupled threshold 
oscillators of relaxation) is laid out that allows epileptology to adopt 
approaches with potential usefulness from the more mature fields of 
seismology and finance. Among those approaches briefly presented in this 
chapter, self-excited marked point processes are, in our opinion, worthy of 
investigation.

The perspectives we provide in this chapter and the approaches we propose
are intended to stimulate new research directions to increase the knowledge
of epilepsy dynamics and, with it, the likelihood of predicting seizures in a
manner that improves quality of life of those to which they afflict.

\section{Glossary}
\begin{itemize}
\item[Coarse-graining]  The procedure that removes ``uninformative'' (for the
task at hand) degrees of freedom to obtain a description
of a system at a more integrated and computationally manageable level. 
Coarse-graining provides a range of techniques
to bridge the ``gap'' between the microscopic and macroscopic levels.
\item[Renormalization group] The meta-theory developed by L. Kadanoff, M. Fisher, K. Wilson and many others,
which allows to construct macroscopic theories of critical phenomena 
at the macro-level, from the 
knowledge of constituents and interactions at the microscopic level. 
\item[Bifurcation]  The phenomenon in which a change in a so-called ``order 
parameter'' causes a qualitative change (from one regime to another) in the 
systems' dynamical behaviors.  The theory of bifurcations has led to a 
classification of  regime changes, which turn out to be reducible
to a limited number of cases. The mathematical description of a
bifurcation is called a normal form, which is a differential equation 
representing the time evolution of the order parameter, given the value(s)
 of the control parameter(s).
\item[Self-organized criticality] A concept introduced in 1987 by P. Bak, C. Tang and K. Wiesenfeld, 
according to which many out-of-equilibrium dynamical spatio-temporal systems, which are slowly driven 
and which exhibit threshold-like responses, tend to self-organize to a dynamical state characterized
by a broad range of avalanche sizes quantified by a power law distribution. A sub-class
of self-organized critical systems can be shown to be made of systems functioning at or close to a standard
critical point (in the sense of phase transitions in statistical physics). It is the non-standard type of 
slow driving of the ``order parameters''  that leads to the attraction of the dynamics to the usually unstable critical point.
\item[Point process] In the field of stochastic processes, one must distinguish 
between two broad classes: (i) Continuous or discrete time processes and (ii)
point processes. An example of the former class is for instance the so-called 
random walk (or Wiener process in mathematical parlance).  
Point processes generate events that are distinct from ``background'' activity. 
 In other words, the value of a point process is zero, except when ``the event''
occurs. In contrast, in the first class of processes, the activity is present at all 
time or occurs in continuing steps. Point processes, also called shot-noise in 
physics and jump processes in finance, are thus particularly suitable to 
describe and model system dynamics characterized by the occurrence of 
events, such as earthquakes, financial crashes and epileptic seizures. 
\end{itemize}

\bibliographystyle{plain}
\bibliography{biblio}

\end{document}